\documentclass{emulateapj}

\usepackage{natbib}
\usepackage{amsmath}

\newcommand{\jhk}{\protect\hbox{$J\!H\!K$}}
\newcommand{\yjhk}{\protect\hbox{$Y\!J\!H\!K$}}
\newcommand{\izyjhk}{\protect\hbox{$izY\!J\!H\!K$}}
\newcommand{\kms}{\,km\,s$^{-1}$}
\newcommand{\dm}{$\Delta m_{15} (B)$}
\newcommand{\um}{$\mu$m}
\newcommand{\mgii}{\ion{Mg}{2}}
\newcommand{\ci}{\ion{C}{1}}
\newcommand{\cii}{\ion{C}{2}}
\newcommand{\lam}{$\lambda$}

\shorttitle{NIR spectroscopy of SN~2011fe}
\shortauthors{Hsiao et~al.}

\begin{document}

\title{The Earliest Near-infrared Time-series Spectroscopy of a Type I\lowercase{a} Supernova}

\def\lco{1}
\def\cfa{2}
\def\car{3}
\def\gem{4}
\def\mpa{5}
\def\qub{6}
\def\fsu{7}
\def\snu{8}
\def\ccc{9}
\def\ucb{10}
\def\uab{11}
\def\ist{12}
\def\aar{13}
\def\a&m{14}

\author{
{E.~Y.~Hsiao}\altaffilmark{\lco},
{G.~H.~Marion}\altaffilmark{\cfa},
{M.~M.~Phillips}\altaffilmark{\lco},
{C.~R.~Burns}\altaffilmark{\car},
{C.~Winge}\altaffilmark{\gem},
{N.~Morrell}\altaffilmark{\lco},
{C.~Contreras}\altaffilmark{\lco},
{W.~L.~Freedman}\altaffilmark{\car},
{M.~Kromer}\altaffilmark{\mpa},
{E.~E.~E.~Gall}\altaffilmark{\mpa,\qub},
{C.~L.~Gerardy}\altaffilmark{\fsu},
{P.~H\"{o}flich}\altaffilmark{\fsu},
{M.~Im}\altaffilmark{\snu},
{Y.~Jeon}\altaffilmark{\snu},
{R.~P.~Kirshner}\altaffilmark{\cfa},
{P.~E.~Nugent}\altaffilmark{\ccc,\ucb},
{S.~E.~Persson}\altaffilmark{\car},
{G.~Pignata}\altaffilmark{\uab},
{M.~Roth}\altaffilmark{\lco},
{V.~Stanishev}\altaffilmark{\ist},
{M.~Stritzinger}\altaffilmark{\aar},
{N.~B.~Suntzeff}\altaffilmark{\a&m}
}
\email{hsiao@lco.cl}

\altaffiltext{\lco}{Carnegie Observatories, Las Campanas Observatory, Colina El Pino, Casilla 601, Chile}
\altaffiltext{\cfa}{Harvard-Smithsonian Center for Astrophysics, 60 Garden Street, Cambridge, MA 02138, USA}
\altaffiltext{\car}{Carnegie Observatories, 813 Santa Barbara St, Pasadena, CA 91101, USA}
\altaffiltext{\gem}{Gemini South Observatory, c/o AURA Inc., Casilla 603, La Serena, Chile}
\altaffiltext{\mpa}{Max-Planck-Institut f\"{u}r Astrophysik, Karl-Schwarzschild-Str. 1, 85741 Garching bei M\"{u}nchen, Germany}
\altaffiltext{\qub}{Astrophysics Research Centre, School of Maths and Physics, Queen’s University Belfast, Belfast BT7 1NN, UK}
\altaffiltext{\fsu}{Florida State University, Tallahassee, FL 32306, USA}
\altaffiltext{\snu}{CEOU/Astronomy Program, Department of Physics \& Astronomy, Seoul National University, Seoul, Korea}
\altaffiltext{\ccc}{Computational Cosmology Center, Computational Research Division, Lawrence Berkeley National Laboratory, 1 Cyclotron Road MS 50B-4206, Berkeley, CA 94611, USA}
\altaffiltext{\ucb}{Department of Astronomy, University of California, Berkeley, CA, 94720-3411, USA}
\altaffiltext{\uab}{Departamento de Ciencias Fisicas, Universidad Andres Bello, Avda. Republica 252, Santiago, Chile}
\altaffiltext{\ist}{CENTRA - Centro Multidisciplinar de Astrof\'isica, Instituto Superior T\'ecnico, Av. Rovisco Pais 1, 1049-001 Lisbon, Portugal}
\altaffiltext{\aar}{Department of Physics and Astronomy, Aarhus University, Ny Munkegade, DK-8000 Aarhus C, Denmark}
\altaffiltext{\a&m}{Physics Department, Texas A\&M University, College Station, TX 77843, USA}

%%%%%%%%%%%%%%
%% Abstract %%
%%%%%%%%%%%%%%

\begin{abstract}

We present ten medium-resolution, high signal-to-noise ratio
near-infrared (NIR) spectra of SN~2011fe from SpeX on the NASA
Infrared Telescope Facility (IRTF) and Gemini Near-Infrared
Spectrograph (GNIRS) on Gemini North, obtained as part of the Carnegie
Supernova Project.  This data set constitutes the earliest time-series
NIR spectroscopy of a Type Ia supernova (SN~Ia), with the first
spectrum obtained at 2.58 days past the explosion and covering $-14.6$
to $+17.3$ days relative to $B$-band maximum.  \ci\ \lam1.0693 \um\ is
detected in SN~2011fe with increasing strength up to maximum
light. The delay in the onset of the NIR \ci\ line demonstrates its
potential to be an effective tracer of unprocessed material.  For the
first time in a SN~Ia, the early rapid decline of the
\mgii\ \lam1.0927 \um\ velocity was observed, and the subsequent
velocity is remarkably constant.  The \mgii\ velocity during this
constant phase locates the inner edge of carbon burning and probes the
conditions under which the transition from deflagration to detonation
occurs.  We show that the \mgii\ velocity does not correlate with the
optical light-curve decline rate \dm.  The prominent break at
$\sim1.5$ \um\ is the main source of concern for NIR k-correction
calculations.  We demonstrate here that the feature has a uniform time
evolution among SNe~Ia, with the flux ratio across the break strongly
correlated with \dm.  The predictability of the strength and the onset
of this feature suggests that the associated k-correction
uncertainties can be minimized with improved spectral templates.

\end{abstract}


%%%%%%%%%%%%%%%%%%
%% Introduction %%
%%%%%%%%%%%%%%%%%%

\section{Introduction}
\label{s:intro}

%SN Ia cosmology

SNe~Ia provide a direct measure of the expansion history of the
universe and have led to the discovery of the accelerated expansion
\citep{1998AJ....116.1009R, 1999ApJ...517..565P}.  The unknown cause
of the accelerated expansion is commonly referred to as ``dark
energy''. Although SNe~Ia remain the most proven technique for
studying dark energy, there are legitimate concerns that systematic
errors coming from the astrophysics of SNe~Ia will ultimately limit
their accuracy.

%cosmology in the NIR

Fortunately, observations in the NIR offer a way forward.
\citet{1985ApJ...296..379E} presented the first SN~Ia NIR Hubble
diagram with a dispersion of only 0.2 mag.  Subsequent studies
indicated that NIR peak luminosity indeed show a smaller intrinsic
scatter than that in the optical \citep[e.g.,][]{2004ApJ...602L..81K,
  2008ApJ...689..377W, 2010AJ....139..120F, 2012MNRAS.425.1007B}, a
result which was supported by the theoretical work of
\citet{2006ApJ...649..939K}.  As opposed to the width-luminosity
relation in the optical \citep{1993ApJ...413L.105P}, NIR peak
luminosities for all except the fastest-declining events have only a
slight dependence on the light-curve shape \citep{2009AJ....138.1584K,
  2012PASP..124..114K}.  Further analyses confirmed that NIR
luminosities without light-curve shape corrections are more effective
distance indicators than corrected optical luminosities
\citep{2009ApJ...704..629M, 2011ApJ...731..120M}.  A key ingredient to
realize the full potential of NIR SN~Ia cosmology is NIR spectroscopy,
such that the peak luminosities can be accurately k-corrected to the
rest frame.  With the limited scope and size of the world's current
NIR spectroscopic sample, the time evolution and the diversity of the
spectral features are to date poorly understood.  These uncertainties
directly affect the determination of the NIR rest-frame peak
luminosities.

%physical diagnostics from NIR spectroscopy

Given that SNe~Ia play a critical role in observational cosmology, it
is as critical as ever to understand the explosion mechanisms and the
progenitor systems.  NIR spectra also provide several key diagnostics
of SN~Ia physics.  The \ci\ \lam1.0693 \um\ line is strong and
isolated in the NIR and is an ideal tracer of primordial,
unprocessed material from the carbon-oxygen white dwarf
\citep{2006ApJ...645.1392M}.  The amount and the distribution of
unprocessed material are key discriminators for SN~Ia explosion
models.  Magnesium is a product of explosive carbon burning and not
oxygen burning.  The strong NIR \mgii\ \lam1.0927 \um\ line provides a
direct measure of the inner boundary of carbon burning and probes the
conditions under which transition from deflagration to detonation
occurs \citep{1998ApJ...496..908W}.  After maximum light, the dramatic
drop in electron scattering opacity, coupled with heavy line
blanketing from fully-processed iron-peak material, rapidly forms the
prominent feature at $1.5-1.9$ \um\ and provides a temperature probe
of the line blanketing material \citep{1998ApJ...496..908W}.

%state of current NIR spectroscopic data

With the advent of NIR narrow-gap semiconductor detectors, pioneering
studies of SN~Ia NIR spectroscopy began in the late 1980s
\citep[e.g.,][]{1987ApJ...315L.129F, 1990AJ....100..223L,
  1992MNRAS.258P..53S, 1994MNRAS.266L..17S}.
\citet{1997MNRAS.290..663B} published the first multi-object sample,
which consisted of NIR spectra taken in late and nebular phases.  The
current sample of SN~Ia NIR spectroscopy largely consists of data of a
few nearby SNe~Ia.  We list here the spectra included in our analysis:
SNe~1994D \citep{1996MNRAS.281..263M}, 1998bu
\citep{1999ApJS..125...73J,2000MNRAS.319..223H,2002AJ....124..417H},
1999by \citep{2002ApJ...568..791H}, 1999ee
\citep{2002AJ....124..417H}, 2002bo \citep{2004MNRAS.348..261B},
2002dj \citep{2008MNRAS.388..971P}, 2003du
\citep{2007A&A...469..645S}, 2004S \citep{2007AJ....133...58K}, 2005cf
\citep{2012MNRAS.427..994G}, 2011iv \citep{2012ApJ...753L...5F}.
\citet{2003ApJ...591..316M, 2009AJ....138..727M} published a catalog
of 41 SN~Ia near-infrared spectra, nearly doubling the world's sample.
Owing to these works, we can begin to examine the statistical
spectroscopic properties of SNe~Ia in the NIR.

%time-series data

Despite the progress made thus far, the NIR spectroscopic sample of
SNe~Ia is not yet ideal.  The NIR sample size is two orders of
magnitude smaller than that in the optical.  Many of the SNe~Ia
observed lack accompanying photometric information.  Most importantly,
there has only been a handful of time-series observations.
Densely-sampled time-series observations are required to understand
how the features evolve with time and vary among object-to-object.
The discovery of the nearby SN~2011fe made possible such observations.

%SN 2011fe

On August 24, 2011, SN~2011fe was detected within hours of its
explosion in M101 \citep{2011Natur.480..344N}.  Its proximity and
early detection provided a unique opportunity to make exquisitely
detailed observations of a SN~Ia.  SN~2011fe has reportedly
representative properties of SNe~Ia and serves as an ideal baseline to
compare to other objects.  It is rapidly becoming the best-studied
SN~Ia to date, with numerous studies from X-ray to radio already
published \citep[e.g.,][]{2011Natur.480..348L, 2011arXiv1112.0247P,
  2012ApJ...744L..17B, 2012ApJ...746...21H, 2012ApJ...750L..19R,
  2012ApJ...750..164C, 2012ApJ...751..134M, 2012ApJ...752L..26P,
  2012ApJ...753...22B, 2012ApJ...754...19M, 2012A&A...546A..12V}.
Early light curves and pre-explosion images of SN~2011fe placed strong
constraints on the progenitor system, ruling out a massive companion
star, disfavoring most theoretical double-degenerate progenitor
systems and excluding Roche-Lobe overflowing red giant and
main-sequence companions to high significance
\citep{2011Natur.480..344N, 2011Natur.480..348L, 2012ApJ...744L..17B};
although, a M dwarf companion may evade these constraints
\citep{2012ApJ...758..123W}.  Ten NIR spectra of SN~2011fe were
obtained in the span of a month, as part of a joint Carnegie Supernova
Project (CSP)-CfA Supernova Group program to obtain a statistical
sample of NIR spectroscopic observations of supernovae (Phillips et
al. in prep).  The SN~2011fe data set constitutes the earliest and
highest signal-to-noise ratio time-series NIR spectroscopy of a SN~Ia.
In this paper, we present several analyses and insights on time
evolution, which would not have been possible without this
one-of-a-kind data set.

%%%%%%%%%%%%%%%%%%
%% Observations %%
%%%%%%%%%%%%%%%%%%

\section{Observation and Reduction}
\label{s:obs}

%SpeX and GNIRS observations

Only 2.58 days past the explosion, the first NIR spectrum of SN~2011fe
was obtained with the SpeX spectrograph \citep{2003PASP..115..362R} on
IRTF.  Nine subsequent NIR spectra were obtained with a regular
three-day cadence with GNIRS \citep{1998SPIE.3354..555E} on Gemini
North.  Relatively short per-frame exposure times were chosen to
prevent the saturation of the telluric OH lines and the supernova.
The object was nodded along the slit using the classical ABBA
technique.  Two ABBA sequences (8 exposures) were obtained at each
epoch.  To correct the science spectra for the effect of telluric
atmospheric absorption, an A0V star was observed at a similar air mass
to that of the science observation.  The slit was oriented along the
parallactic angle for all observations \citep{1982PASP...94..715F}.
The spectra from both SpeX and GNIRS presented here have similar
medium spectral resolving power of $R\sim1000-2000$. With the
exception of the first SpeX spectrum, all of the resulting spectra
have very high signal-to-noise ratio of well above 100.  A journal of
the spectroscopic observations is given in Table~\ref{t:obs}.

\begin{deluxetable*}{cccccccccc}[h!]
\centering
\tablecolumns{10}
\tablecaption{List of Observations\label{t:obs}}
\tablehead{
UT   & UT  &            & Phase wrt & Phase wrt   & Number of & Integration & SN 2011fe & Telluric & Telluric \\
Date & MJD & Instrument & Explosion & $B$ Maximum & Exposures & Time        & Air mass   & Standard & Air mass }
\startdata
2011-08-26 & 55799.27 &   SpeX &   2.58 d & $-$14.6 d &   8 &  1200 s &  2.00 & HD143187 &  1.92 \\
2011-08-28 & 55801.25 &  GNIRS &   4.56 d & $-$12.6 d &   8 &  2400 s &  1.78 & HIP69366 &  1.91 \\
2011-08-31 & 55804.23 &  GNIRS &   7.55 d & $-$9.7 d  &   8 &   960 s &  1.77 & HIP69366 &  1.77 \\
2011-09-03 & 55807.24 &  GNIRS &  10.55 d & $-$6.7 d  &   8 &  1200 s &  1.89 & HIP69366 &  2.05 \\
2011-09-07 & 55811.11 &  GNIRS &  14.42 d & $-$2.8 d  &   8 &   960 s &  1.88 & HIP69366 &  1.99 \\
2011-09-10 & 55814.21 &  GNIRS &  17.52 d & $+$0.3 d  &   8 &   960 s &  1.87 & HIP69366 &  1.95 \\
2011-09-13 & 55817.24 &  GNIRS &  20.55 d & $+$3.3 d  &   8 &   960 s &  2.28 & HIP69366 &  2.62 \\
2011-09-18 & 55822.13 &  GNIRS &  25.45 d & $+$8.2 d  &   8 &   700 s &  2.25 & HIP69366 &  2.54 \\
2011-09-22 & 55826.22 &  GNIRS &  29.53 d & $+$12.3 d &   8 &   960 s &  2.35 & HIP69366 &  2.84 \\
2011-09-27 & 55831.21 &  GNIRS &  34.52 d & $+$17.3 d &   8 &   960 s &  2.54 & HIP72520 &  2.35
\enddata
\tablecomments{The time of explosion of MJD 55796.687
  \citep{2011Natur.480..344N} and the time of $B$-band maximum of MJD
  55813.9 \citep{2012A&A...546A..12V} were adopted.}
\end{deluxetable*}

%SpeX reduction

The SpeX spectrum was obtained in the cross-dispersed mode, utilizing
a grating and prism cross-dispersers, with a 0$\farcs$5 slit.  This
configuration yields a wavelength coverage from 0.8 to 2.5
\um\ divided over six orders, with gaps over the regions of strong
telluric absorptions between the \jhk\ bands.  With the choice of the
0$\farcs$5 slit width, the resulting spectral resolution is in the
range of $\sim 1600-2000$.  The data were calibrated and reduced using
the IDL pipeline \texttt{Spextool} \citep{2004PASP..116..362C},
specifically designed for the reduction of SpeX data.  The pipeline
performed steps of pair subtraction, flat-fielding, aperture
definition, spectral tracing and extraction, and wavelength
calibration.  The 1D spectra from separate exposures were then
combined.  Corrections for telluric absorption were performed using
the IDL tool \texttt{xtellcor} developed by
\citet{2003PASP..115..389V}.  To construct a telluric correction
spectrum free of stellar absorption features, a model spectrum of Vega
was used to match and remove the hydrogen lines of Paschen and
Brackett series from the AV0 telluric standard.  The resulting
telluric correction spectrum was also used for the absolute flux
calibration, as the NIR continua of A0V stars are reasonably
approximated by a blackbody, with the temperature (typically
$\sim$10,000 K) determined by the color of the star.

%GNIRS reduction

The GNIRS spectra were observed in the cross-dispersed mode, in
combination with the short-wavelength camera, a 32 lines per mm
grating and a 1$\farcs$0 slit.  This configuration allows for a wide
continuous wavelength coverage from 0.8 to 2.5 \um, divided over six
orders.  Because of the choice of the 1$\farcs$0 slit width, the
resulting spectral resolution is lower than the SpeX spectrum and in
the range of $\sim 1000-1300$.  The GNIRS data were calibrated and
reduced using the \texttt{gnirs} pipeline within the Gemini
IRAF\footnote{IRAF is distributed by the National Optical Astronomy
  Observatory, which is operated by the Association of Universities
  for Research in Astronomy (AURA) under cooperative agreement with
  the National Science Foundation.} package, specifically developed
for the reduction of GNIRS data.  The steps began with non-linearity
correction, locating the spectral orders and flat-fielding.  Sky
subtractions were performed for each AB pair closest in time, then the
2D spectra were stacked.  Spatial distortion correction and wavelength
calibrations were applied before the 1D spectrum was extracted.  To
perform telluric correction, the stellar hydrogen lines were first
removed from the telluric star spectrum.  The IRAF task
\texttt{telluric} was then used to interactively adjust the relative
wavelength shift and flux scale to divide out telluric features
present in the science spectrum.  A blackbody spectrum is then assumed
for the telluric star for the flux calibration.  Some high-frequency
pattern noise with pixel-wide correlations were evident in isolated
regions of the spectrum; this is a known issue for the GNIRS detector.
These were removed with a low-pass filter.  There were adequate
overlaps between orders to join the spectra; however, some overlaps
occurred in the uncertain regions of strong telluric absorptions
between the \jhk\ bands.  We used the \jhk\ photometry of
\citet{2012ApJ...754...19M} as a guide to scale the relative flux
before joining the spectral orders.

%%%%%%%%%%%%%
%% Spectra %%
%%%%%%%%%%%%%

\section{NIR Spectra of SN~2011\lowercase{fe}}
\label{s:spec}

%the NIR spectra

\begin{figure*}
\begin{center}
\epsscale{1.2}
\plotone{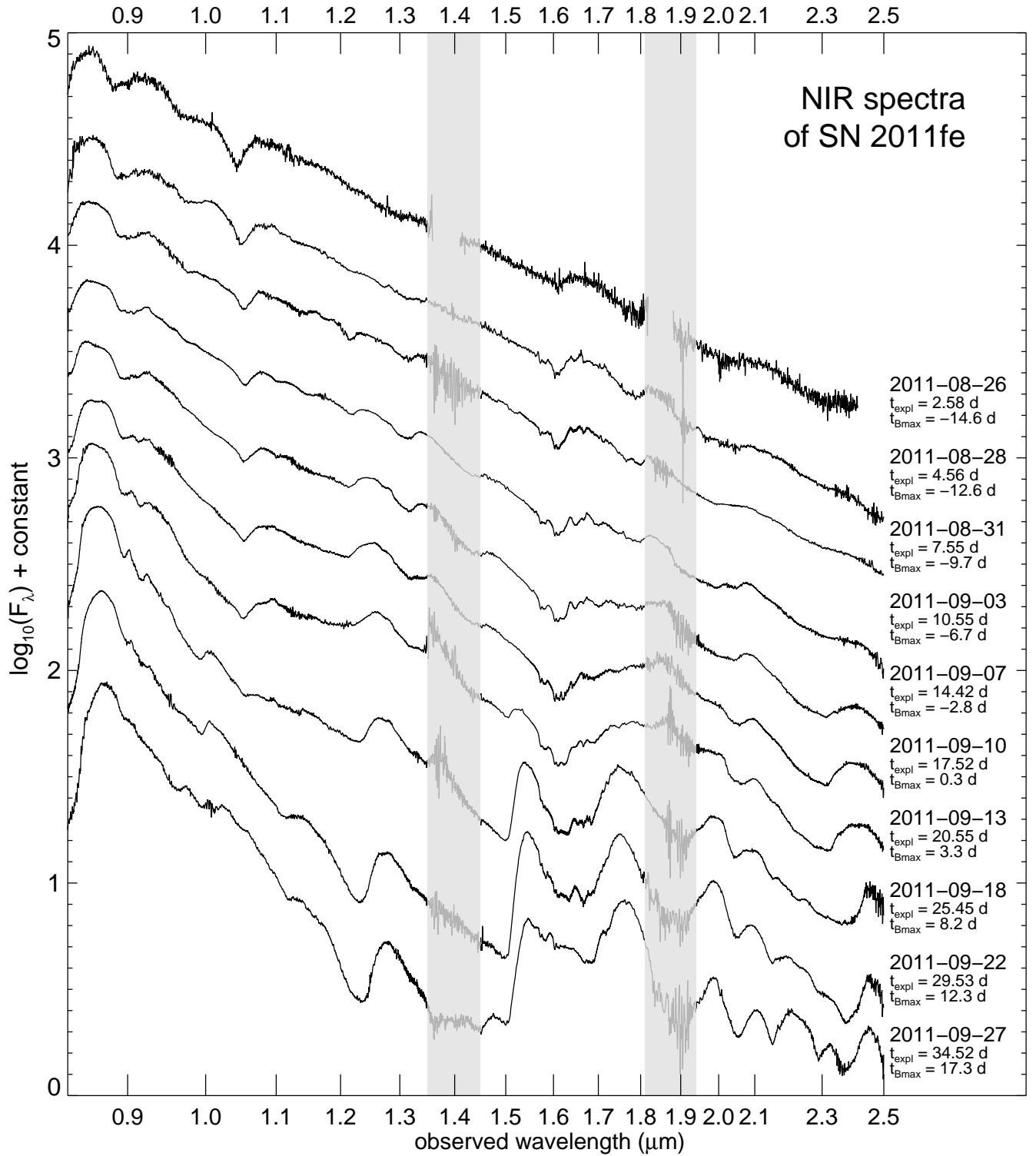}
\caption{NIR spectra of SN~2011fe.  The wavelength is as observed.
  The logarithmic scale of flux is plotted.  For clarity, the spectra
  are scaled such that they have constant spacing at 1 \um.  The UT
  date of observation, phase relative to explosion and phase relative
  to $B$-band maximum are labeled for each spectrum.  Grey vertical
  bands mark the regions of the strongest telluric absorptions.}
\label{f:spec}
\end{center}
\end{figure*}

Ten NIR spectra of SN~2011fe are presented in
Figure~\ref{f:spec} \footnote{The spectra can be downloaded at
  http://csp2.lco.cl/hsiao/}.  The early NIR spectra are dominated by
electron scattering with the continua resembling the Rayleigh-Jeans
tail of a blackbody.  The spectral slope remains quite constant until
around maximum light.  As the ejecta expands and electron scattering
opacity decreases, the continuum drops, and a well-defined photosphere
no longer exists.  Just past maximum light and at regions of heavy
line blanketing by iron-peak elements, the lines are formed at larger
effective radii producing features seemingly in emission
\citep{1998ApJ...496..908W}.  The combination of these lines in
``emission,'' and the decreased continuum, forms the most prominent
feature in the NIR between $1.5-1.9$ \um.

%fluorescence

Alternatively, the origin of the increased flux at $1.5-1.9$ \um\ was
shown to be caused by increased fluorescence in the region
\citep{2012MNRAS.427..994G}. Line blanketing of iron-group elements is
very efficient in blocking flux in the ultraviolet (UV) part of the
spectrum. In contrast the opacity in the optical and NIR is
significantly lower \citep[e.g.,][]{2000ApJ...530..757P}. Thus photons
absorbed in the UV and redistributed to the NIR by line fluorescence
may escape from the ejecta \citep{2009MNRAS.398.1809K}. At wavelengths
where line fluorescence is highly effective, this can lead to an
increased flux.

%synapps fits of early spectra

\begin{figure*}
\begin{center}
\epsscale{1.2}
\plotone{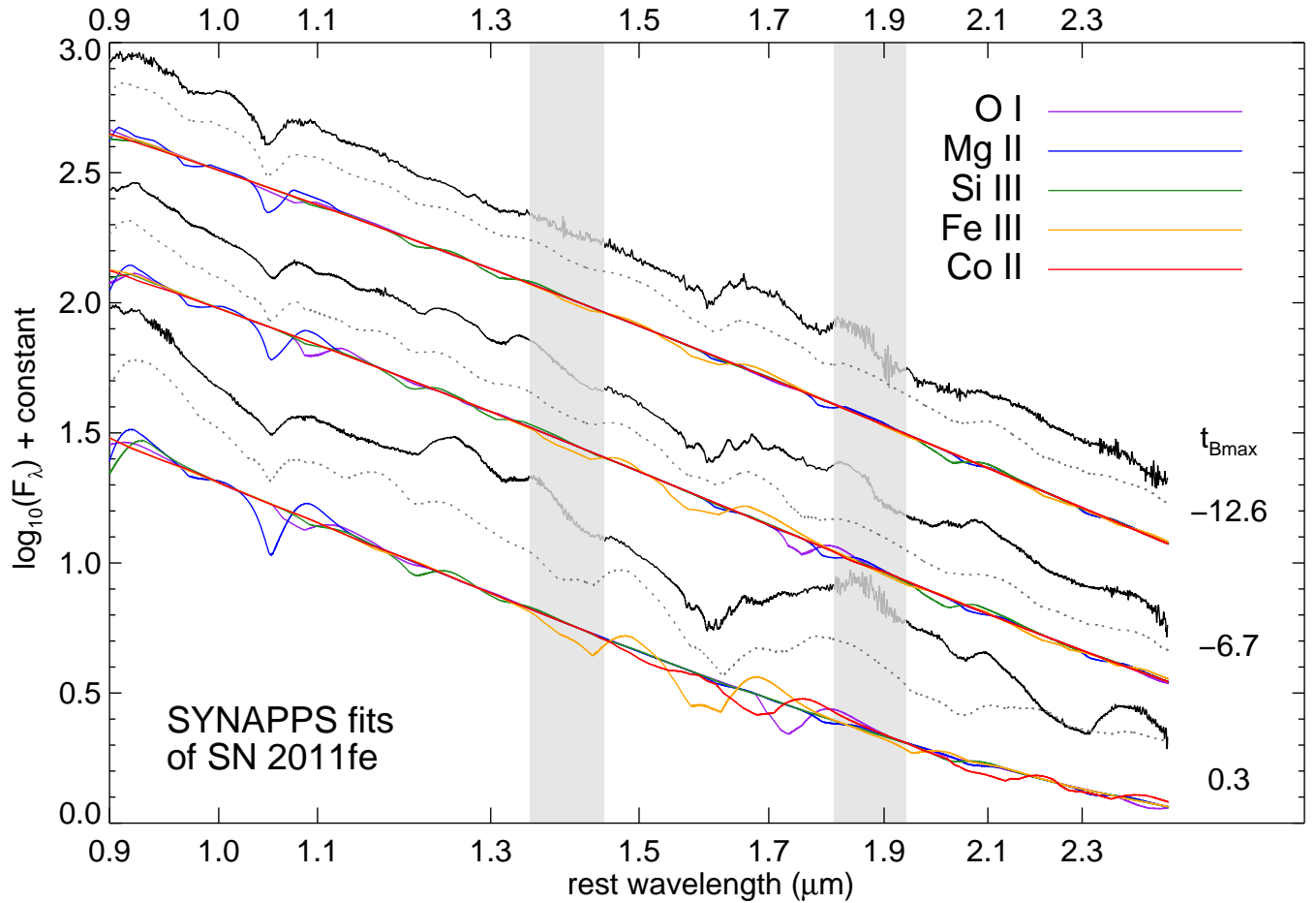}
\caption{\texttt{SYNAPPS} fits for three early NIR spectra of
  SN~2011fe.  The observed spectra and the \texttt{SYNAPPS} fits are
  plotted in black solid and grey dotted curves, respectively.  The
  colored curves show individual contributions from the most prominent
  ions.  The wavelength axis is plotted in the rest frame of the host
  galaxy.  The phase relative to $B$-band maximum is labeled for each
  spectrum.}
\label{f:synapps}
\end{center}
\end{figure*}

We fit three early NIR spectra with the automated spectrum synthesis
code \texttt{SYNAPPS} \citep{2011PASP..123..237T}, derived from
\texttt{SYNOW} \citep{2005PASP..117..545B}.  The results are shown in
Figure~\ref{f:synapps}.  \texttt{SYNAPPS} uses a highly parameterized,
but fast spectrum synthesis technique, useful for identifying the ions
that form the observed features.  We included in the fits ions which
are commonly identified or expected to be present in normal SN Ia
optical and NIR spectra: \ion{C}{1}, \ion{C}{2}, \ion{O}{1},
\ion{Mg}{2}, \ion{Si}{2}, \ion{Si}{3}, \ion{S}{2}, \ion{Ca}{2},
\ion{Fe}{2}, \ion{Fe}{3}, \ion{Co}{2} and \ion{Ni}{2}.  Such an
expansive list of ions was included, such that \texttt{SYNAPPS} can
objectively determine the absence or presence of each ion species.
The most prominent features at this epoch are attributed to
intermediate-mass species, such as \ion{O}{1}, \ion{Mg}{2} and
\ion{Si}{3}.  The strong \mgii\ \lam1.0927 \um\ absorption is formed
at a relatively isolated region near 1.05 \um.  Signatures of unburned
\ion{C}{1} were detected and will be discussed in detail in
Section~\ref{s:carbon}.  Redward of the strong \mgii\ line near 1.05
\um, four absorption features from 1.1 \um\ to 1.4 \um\ were
identified as \ion{O}{1}/\ion{Si}{3}, \ion{C}{1}/\ion{Ca}{2},
\ion{Si}{3}, and \ion{Si}{3}, respectively.

%iron-peak products

In terms of iron-peak elements, \texttt{SYNAPPS} interprets the strong
feature near 1.6 \um\ as \ion{Fe}{3}.  This feature was identified as
\ion{Mg}{2}/\ion{Si}{2}/\ion{Co}{2} by \citet{2009AJ....138..727M} and
\ion{Si}{2} by \citet{1998ApJ...496..908W} and
\citet{2012MNRAS.427..994G}.  Two absorption minima near 1.6 \um\ were
resolved in SN~2011fe. Both minima are attributed to \ion{Fe}{3} with
weaker influences from \ion{Mg}{2}, \ion{Si}{2} and \ion{Si}{3} in the
redder of the two absorptions.  The \ion{Fe}{3} interpretation also
yielded good fits for several other features in and around the H band
(Figure~\ref{f:synapps}). \ion{Ni}{2} appears to be weak in all three
early spectra, while \ion{Co}{2} increases in strength with time.  The
increasing influence of \ion{Co}{2} is attributed to both decreasing
opacity and increasing abundance of $^{56}$Co from the radioactive
decay of $^{56}$Ni.

%ionization condition

The presence or absence of a certain ionization stage of a species is
regulated by the combination of temperature and the strength of the
lines in question.  While \cii\ was positively identified in the
optical spectra of SN~2011fe \citep{2012ApJ...752L..26P}, the
\cii\ lines are weak in the NIR. Instead, \ci\ is identified in the
NIR, with the strongest influence near maximum light rather than at
earlier phases (Section~\ref{s:carbon}).  While \ion{Si}{2} is the
hallmark feature of a SN~Ia in the optical, \ion{Si}{3} was found to
have more identifiable features in the NIR spectra of SN~2011fe.  At
early phases, \ion{Fe}{3} has stronger influences than \ion{Fe}{2} in
the NIR, similar to the behavior in the optical.

%W7 model

\begin{figure}
\begin{center}
\epsscale{1.2}
\plotone{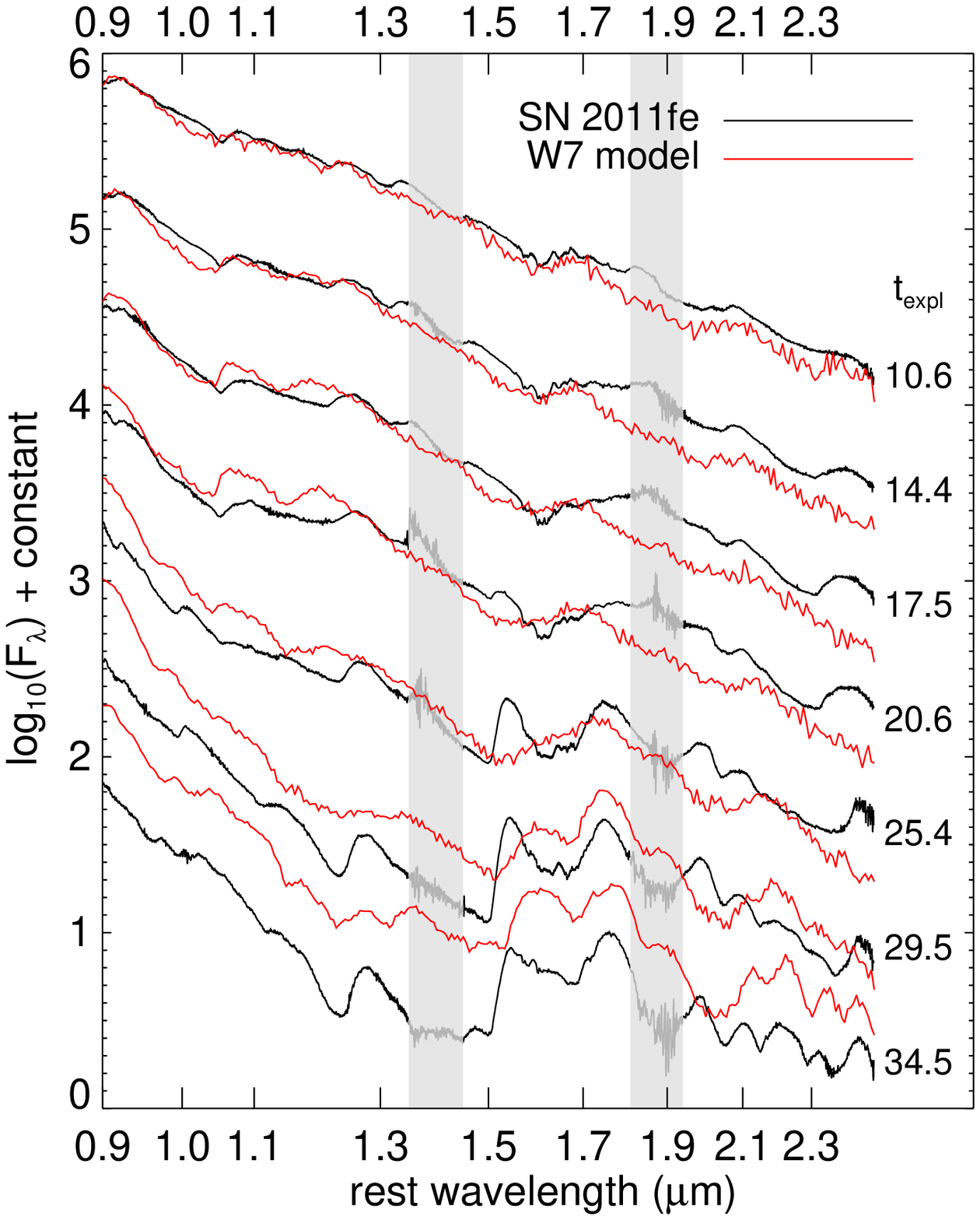}
\caption{Comparison between the synthetic W7 model spectra and the NIR
  spectra of SN~2011fe.  The model spectra were calculated using the
  radiative transfer code \texttt{ARTIS} for the W7 model.  The
  observed and the model spectra are plotted in black and red,
  respectively. The wavelength axis is plotted in the rest frame of
  the host galaxy of SN~2011fe.  The absolute flux scales of the
  observed and model time series were matched at the first comparison
  epoch, then the same offset was applied to each subsequent pair of
  spectra.  The phase relative to explosion is labeled for each
  spectrum.}
\label{f:w7}
\end{center}
\end{figure}

Recently, \citet{2012MNRAS.427..994G} showed the close resemblance
between the NIR spectra of SN~2005cf and the synthetic spectra of the
hydrodynamical explosion model W7 \citep{1984ApJ...286..644N,
  1999ApJS..125..439I}.  They found that in the NIR, the emerging flux
is almost entirely due to fluorescent emission with a few features
formed by strong P-Cygni absorptions.  Here, we make the comparison
with the NIR spectra of SN~2011fe, using the same synthetic spectra
without modifications (Figure~\ref{f:w7}).  While \texttt{SYNAPPS} is
useful for identifying the line forming ion species, it does not
provide detailed post-processing of stellar explosion models.  The W7
model spectra in Figure~\ref{f:w7} were calculated using the Monte
Carlo radiative transfer code \texttt{ARTIS}
\citep{2009MNRAS.398.1809K}. It provides a fully self-consistent
solution to the radiative transfer problem and produces time-dependent
spectral synthesis without any free parameters.  As was found by
\citet{2012MNRAS.427..994G}, the spectral features of the observed and
model spectra show encouraging similarities. A few differences are
noted here. Slight discrepancy is observed in the time evolution of
the NIR brightness. The two spectra of SN~2011fe near a month past
explosion or 10 days after B-band maximum are consistently fainter
than the model spectra.  The strong \mgii\ \lam1.0927 \um\ feature
near 1.05 \um\ is present in the early model spectra; however,
discrepancies in the velocity evolution are evident between the
observed and model spectra.  While the \mgii\ \lam1.0927 \um\ velocity
of SN~2011fe decreases rapidly (Section~\ref{s:magnesium}), the
velocity in the model spectra remains high.  Also, the observed strong
break near 1.5 \um\ (Section~\ref{s:iron}), one of the most prominent
spectral features in the NIR, is not reproduced by the model until
approximately one month past the explosion.  Even then, the velocity
of the feature does not match the observations.

%%%%%%%%%%%%%
%% Physics %%
%%%%%%%%%%%%%

\section{Diagnostics of SN I\lowercase{a} Physics}
\label{s:phys}

%deflagration to detonation transition

There is currently no full theoretical description of the explosion
mechanism of SNe~Ia that accounts for the observed diversity.  In the
most established model, carbon is ignited near the center of a
carbon-oxygen white dwarf, as it approaches the Chandrasekhar mass
\citep{1960ApJ...132..565H}.  The initial thermonuclear carbon burning
starts in a subsonic deflagration \citep{1976Ap&SS..39L..37N}.  To
produce partially-processed, intermediate-mass material observed in
SNe~Ia, the transition to detonation is ``delayed,'' such that the
white dwarf is able to expand appreciably to allow for partial
burning.  A possible transition to a supersonic detonation then
rapidly incinerates the star \citep{1991A&A...245..114K,
  1992ApJ...393L..55Y}.  While some mechanisms for the transition from
deflagration in the carbon flame to the subsequent detonation in the
oxygen flame have been proposed, its nature remains uncertain and
controversial \citep[e.g.,][]{1997ApJ...478..678K,
  1997ApJ...475..740N, 2004ApJ...612L..37P, 2008ApJ...689.1173A,
  2009ApJ...704..255W, 2011ApJ...734...37W}.  The conditions under
which the transition occurs are still essentially treated as free
parameters within models.

%physical diagnostics in NIR spectra

Early NIR spectra provide several powerful diagnostics of the physics
of SN~Ia explosions, and can serve as discriminators for competing
explosion models.  Since magnesium is a product of explosive carbon
burning and not oxygen burning, the NIR \mgii\ lines provide a direct
measure of the inner boundary of carbon burning, and could place
meaningful constraints on the conditions under which the transition to
detonation occurs \citep{1998ApJ...496..908W}.  The location of the
boundary is also particularly sensitive to the transition density,
which is inextricably linked to the amount of $^{56}$Ni produced
\citep[e.g.,][]{1995ApJ...444..831H, 2002ApJ...568..791H}.  The
presence or absence of unprocessed material is one of the key
predictions of SN~Ia explosion models.  The NIR \ci\ lines provide an
independent confirmation of the findings from the intensively-studied
optical \cii\ lines.  Past maximum light, the continuum opacity from
electron scattering decreases dramatically. Coupled with heavy line
blanketing from fully-processed, iron-peak material, the NIR spectrum
probes a large range of depths and the temperature of the line
blanketing material \citep{1998ApJ...496..908W}.  In this section, we
present several techniques to quantify the characteristics of NIR
spectra and discuss the use of these measurements as various
diagnostics of SN~Ia physics.

%%%%%%%%%%%%
%% Carbon %%
%%%%%%%%%%%%

\subsection{Unburned Carbon}
\label{s:carbon}

%explosion models

Since oxygen is also produced from carbon burning, carbon provides the
most direct probe of the primordial material from the progenitor
carbon-oxygen white dwarf.  The quantity, distribution and incidence
of unburned carbon in SNe~Ia provide important constraints for
explosion models.  Turbulent deflagration models predict that a large
amount of unprocessed carbon should be left over
\citep{2003Sci...299...77G, 2007ApJ...668.1132R}.  In general, a
transition to detonation would result in complete carbon burning
\citep{2002ApJ...568..791H, 2006ApJ...645.1392M, 2009Natur.460..869K};
although, exceptions have been found. For the lowest transition
density in a grid of delayed detonation models, both the model and the
observations of SN~1999by showed strong \ci\ signatures
\citep{2002ApJ...568..791H}, but SN~1999by is significantly
subluminous \citep{2000A&A...361...63T} and not representative of a
normal SN Ia.  \citet{2004PhRvL..92u1102G} showed that pockets of cool
and dense unburned material could be pulled down toward the center and
escape the detonation wave.

\begin{figure}
\begin{center}
\epsscale{1.2}
\plotone{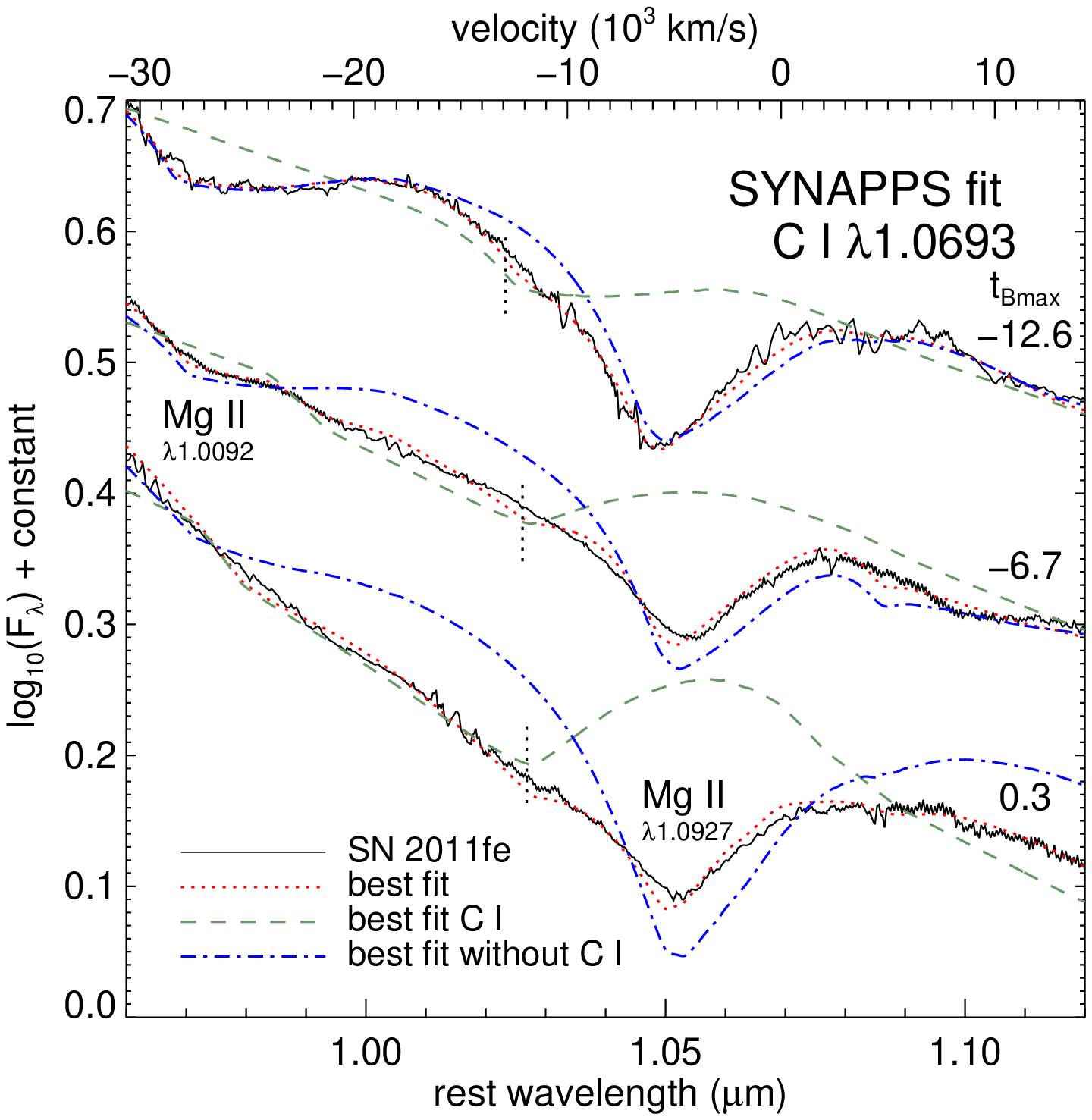}
\caption{\texttt{SYNAPPS} fits of three early NIR spectra of SN~2011fe
  in the region of the \ci\ \lam1.0693 \um\ line.  The velocity axis
  is plotted with respect to the rest wavelength of \ci\ \lam1.0693
  \um.  The spectra are plotted as solid black curves.  The best-fit
  synthesized spectra are plotted as follows: with all ions, with only
  \ci, and with all ions except \ci. These are plotted as red dotted,
  green dashed, and blue dash-dotted curves, respectively.  The
  vertical dotted lines mark the location of the best-fit minimum
  photospheric velocity.  The phases relative to $B$-band maximum
  light are noted.}
\label{f:c_synapps}
\end{center}
\end{figure}

%observations

Unburned carbon in SNe~Ia has been the subject of several intensive
studies in recent years.  Examining the \ci\ lines in NIR spectra of
three normal SNe~Ia without the aid of spectrum synthesis techniques,
\citet{2006ApJ...645.1392M} concluded that the abundance of
unprocessed material is low.  \citet{2007ApJ...654L..53T} presented
the most convincing detection of the \cii\ \lam0.6580 \um\ feature in
the early optical spectra of SN~2006D.  In subsequent studies with
larger optical spectroscopic samples, $20-30$\%\ of the early spectra
were found to show \cii\ signatures \citep{2011ApJ...743...27T,
  2011ApJ...732...30P, 2012ApJ...745...74F, 2012MNRAS.425.1917S}.
Several studies have noted that this fraction represents the lower
limit, as noise, velocity blueshift, and the phase at which the SNe~Ia
were followed up could affect the \cii\ detection
\citep[e.g.,][]{2007PASP..119..709B, 2012ApJ...745...74F}.  The
velocities of the detected \cii\ lines are generally low.  There have
also been hints that SNe~Ia with detected \cii\ have preferentially
bluer colors and narrower light curves \citep{2011ApJ...743...27T,
  2012ApJ...745...74F}, but this finding was not held up in the
examinations of other data sets \citep[e.g.,][]{2012AJ....143..126B,
  2011ApJ...732...30P}.  Nonetheless, there appears to be a consensus
that the mass fraction of the photospheric carbon is low
\citep[e.g.,][]{2003AJ....126.1489B, 2006ApJ...645.1392M,
  2007ApJ...654L..53T, 2008ApJ...677..448T}

\begin{figure}
\begin{center}
\epsscale{1.2}
\plotone{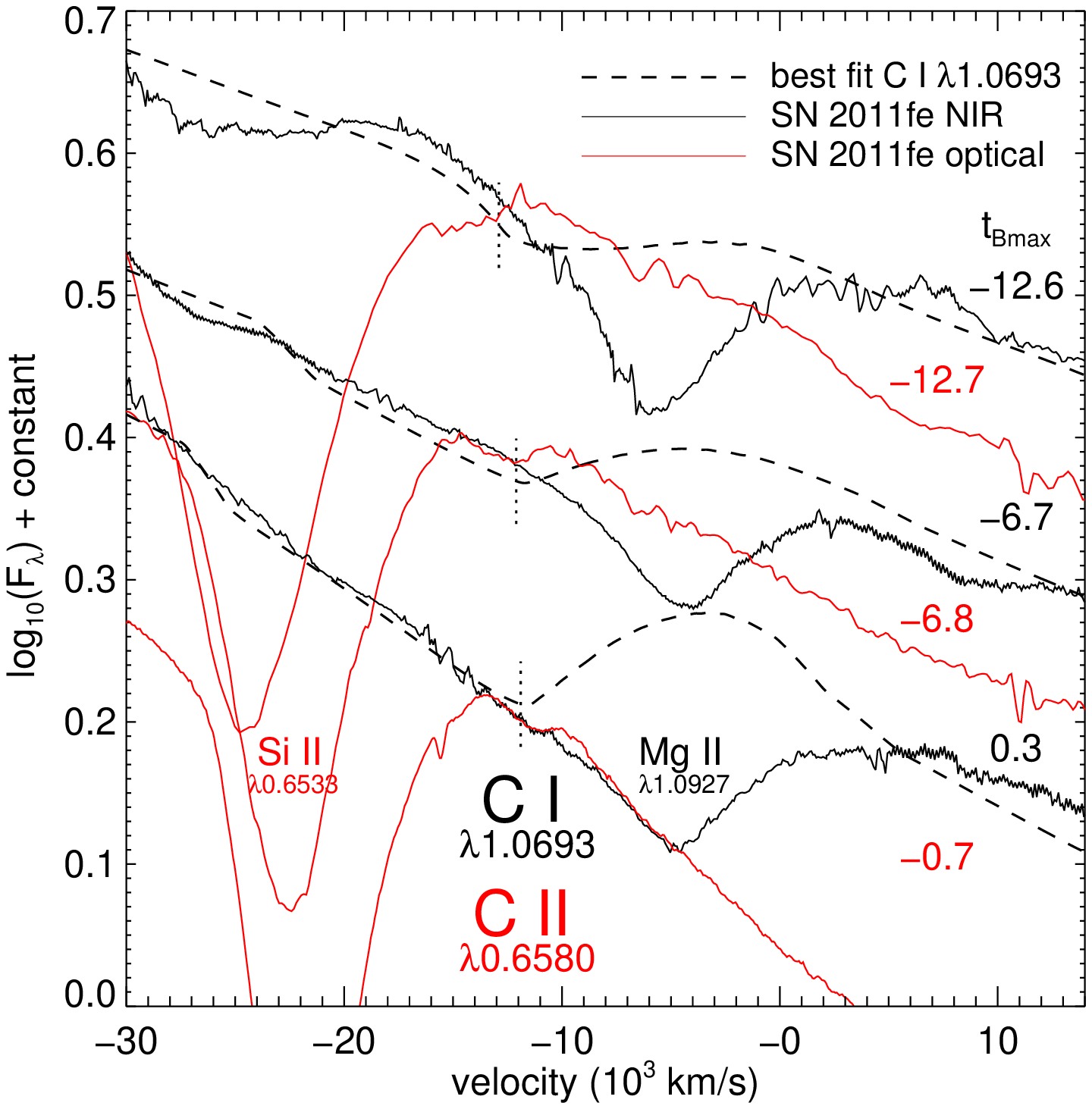}
\caption{Comparison of the NIR \ci\ \lam1.0693 \um\ and the optical
  \cii\ \lam0.6580 \um\ lines of SN~2011fe in velocity space.  The
  velocity axis is plotted with respect to the rest wavelength of the
  \ci\ \lam1.0693 \um\ and \cii\ \lam0.6580 \um\ lines.  The NIR
  spectra are plotted as black curves.  The optical spectra are taken
  from \citet{2012ApJ...752L..26P} and plotted as red curves.  The
  \ci\ profiles from the best-fit synthesized spectra are plotted as
  black dashed curves.  The vertical dotted lines mark the location of
  the best-fit minimum photospheric velocity.  The phases relative to
  $B$-band maximum light for the NIR and optical spectra are labeled
  in black and red, respectively.}
\label{f:c_optical}
\end{center}
\end{figure}

%synapps analysis

The automated spectrum synthesis code \texttt{SYNAPPS} has been
employed successfully to identify the \cii\ \lam0.6580 \um\ lines in
early optical spectra \citep{2007ApJ...654L..53T, 2011ApJ...743...27T,
  2012ApJ...752L..26P}.  Here we concentrate on the study of the
strongest \ci\ line in the NIR at 1.0693 \um\ using \texttt{SYNAPPS}.
The weaker \ci\ lines are included in the fit, but are heavily blended
and more difficult to detect. For example, the
\ci\ \lam\lam0.9087,1.1756 \um\ lines are both blended with
\ion{O}{1}, \ion{Si}{3}, and \ion{Ca}{2} lines.  \ci\ \lam1.0693 \um,
at the largest expected range of velocities, lies in between the
strong \mgii\ \lam1.0927 \um\ line, usually blueshifted to $\sim1.05$
\um, and the weaker \mgii\ \lam1.0092 \um\ line, usually blueshifted
to $\sim0.97$ \um.  The fits were done for three early spectra over
the entire NIR wavelength range (Figure~\ref{f:synapps}).  The results
zoomed in on the region containing the line profiles of
\ci\ \lam1.0693 \um\ are shown in Figure~\ref{f:c_synapps}. The
best-fit synthesized spectra show very close matches to the observed
spectra.  In the first spectrum, at approximately 2 weeks before
maximum, the \ci\ signature appears weak, and was not required to
obtain a decent fit in the region.  Starting from a week before
maximum, the story changes.  The profile begins to take on a flattened
shape (Figure~\ref{f:c_synapps}). The presence of \ci\ is now required
to suppress the emission component of the \mgii\ \lam1.0092
\um\ P-Cygni profile, which is evident at 2 weeks before maximum.
Near maximum light, the profile remains flat, and the influence of
\ci\ is even stronger.  Our spectrum has a high enough signal-to-noise
ratio to resolve the ``notch'' from \ci\ \lam1.0693 \um\ near 1.03
\um. The strengthening of the \ci\ signature with time may be a result
of the increasing ionization fraction of \ci\ versus \cii, as the
ejecta expands and cools.  The unburned carbon in these early NIR
spectra was detected at velocities consistent with the photospheric
velocity of partially and completely synthesized material.  The
blueshift of the \ci\ line was observed at 12,900 \kms\ at 2 weeks
before maximum and leveled off at $\sim$12,000 \kms\ from a week
before maximum light. Note that in past analyses, the flattened
feature has been attributed to an extended wing of \mgii\ \lam1.0927
\um\ \citep[e.g.,][]{2009AJ....138..727M}.  This is produced by a
detonation front which leads to partial burning up to very high
velocities \citep{2002ApJ...568..791H}.

\begin{figure}
\begin{center}
\epsscale{1.2}
\plotone{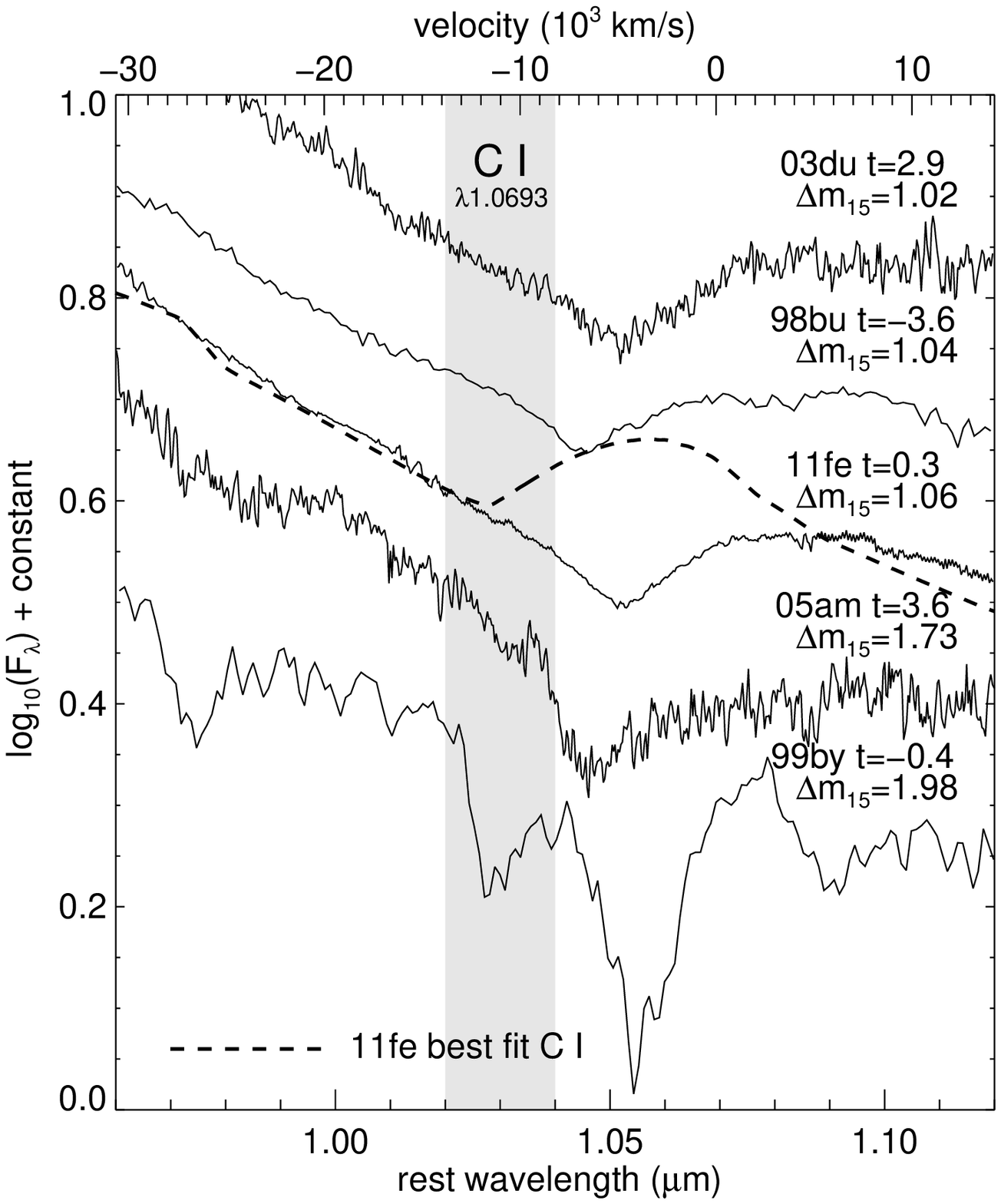}
\caption{Comparison of the NIR \ci\ \lam1.0693 \um\ profile for five
  SNe~Ia near maximum light.  The velocity axis is plotted with
  respect to the rest wavelength of \ci\ \lam1.0693 \um.  The
  \ci\ profile from the best-fit synthesized spectrum of SN~2011fe is
  plotted as a black dashed curve.  The time of $B$-band maximum and
  $B$-band light-curve decline rate \dm\ of the literature spectra
  were taken from \citet{2009ApJ...700..331H}.  The phase relative to
  $B$-band maximum and \dm\ are noted for each spectrum.}
\label{f:c_compare}
\end{center}
\end{figure}

%comparison to optical

Our fits were performed without assuming any initial conditions based
on the results from the optical, such as strength and velocity.  In
Figure~\ref{f:c_optical}, the \ci\ \lam1.0693 \um\ feature from the
NIR and \cii\ \lam0.6580 \um\ feature from the optical of SN~2011fe
are compared in velocity space.  Optical spectra taken at phases
closest to those of the NIR spectra are shown.  The \cii\ features
were positively identified in the early optical spectra of SN~2011fe
and showed decreasing strength with time \citep{2012ApJ...752L..26P}.
The velocities of the \ci\ and \cii\ features match exceptionally
well, indicating that these signatures are from the same line-forming
material.  The interpretation from \texttt{SYNAPPS} that
\ci\ \lam1.0693 \um\ is the source of the flattened profile on the
emission wing of the \mgii\ \lam1.0092 \um\ P-Cygni profile appears to
be correct.

%comparisong to other NIR spectra

It is worth noting here that the flattened profile near 1.03 \um\ is a
common feature for brighter and slower-declining SNe~Ia near maximum
light (Figure~\ref{f:c_compare}).  For fainter, faster-declining
SNe~Ia, the \ci\ signatures appear to be stronger and easily
detectable at maximum light. SN~1999by is the extreme of this example
with conspicuous presence of the \ci\ line.  This effect is
reminiscent of the familiar sequence of the \ion{Si}{2} \lam0.5972
\um\ strength \citep{1995ApJ...455L.147N}.  It is possible to
interpret from these observations that the majority of SNe~Ia, at both
ends of the brightness extreme, harbor some unprocessed carbon deep in
the ejecta.  Such a conclusion is discrepant from the $20-30$\% found
by recent optical studies \citep[e.g.,][]{2011ApJ...743...27T,
  2012ApJ...745...74F, 2012MNRAS.425.1917S}; although, there are
indications that the fraction of detection is higher with earlier
spectra \citep[e.g.,][]{2012ApJ...745...74F}.  Since the finding of
ubiquitous unprocessed carbon would have profound implications for our
understanding of SN~Ia explosions, a larger NIR sample and further
studies are warranted for such a claim.

%C I in the NIR is a better tracer

Nonetheless, we have shown here that the \ci\ \lam1.0693 \um\ feature
in the NIR is potentially a better tracer of unprocessed material than
\cii\ \lam0.6580 \um\ in the optical.  While the signature of
\cii\ \lam0.6580 \um\ is the strongest between one and two weeks
before maximum and fades with time, the influence of \ci\ \lam1.0693
\um\ appears to be the strongest at maximum light in SN~2011fe.  A
change in the ionization condition, as the temperature cools, may have
fortuitously brought about a delay in the onset of the \ci\ feature.
If this behavior can be verified in other SNe~Ia, the problem of the
incompleteness of very early optical spectroscopic samples could be
circumvented by using a sample of NIR spectra near maximum light.
Well situated in a relatively isolated region between the two
\mgii\ lines, the NIR \ci\ line also has the advantage that it is
unblended with other strong absorptions for a large range of
velocities; while \cii\ \lam0.6580 \um\ may be blueshifted into
\ion{Si}{2} \lam0.6355 \um\ at a velocity higher than 15,000
\kms\ \citep{2007PASP..119..709B}.

%%%%%%%%%%%%%%%
%% Magnesium %%
%%%%%%%%%%%%%%%

\subsection{Inner Edge of Magnesium}
\label{s:magnesium}

\begin{figure}
\begin{center}
\epsscale{1.2}
\plotone{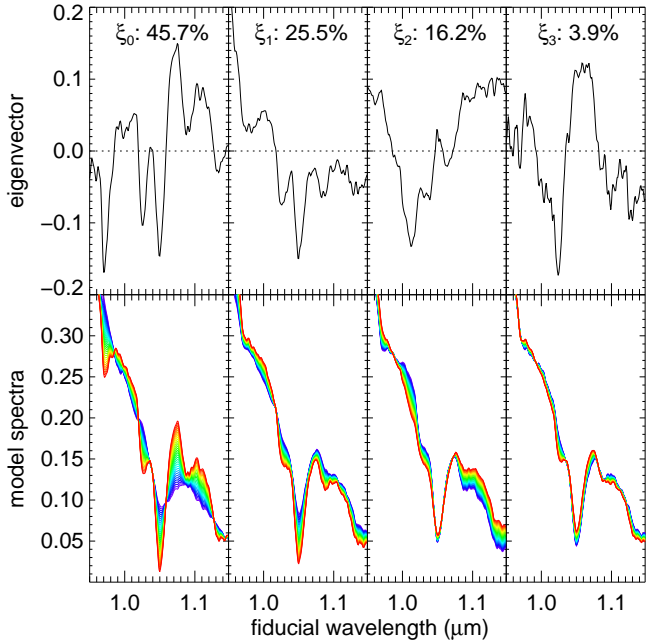}
\caption{Principal component model of the region centered on the
  \mgii\ \lam1.0927 \um\ line.  The first four principal components
  are shown.  For the model spectra, the effects of 1 $\sigma$
  variation in each projection $p_j$ are shown.  The eigenvalue for
  each principal component is noted on the top and is expressed as a
  fractional value, such that the sum of the $N$ eigenvalues totals to
  1.  It is a measure of the amount of variation each principal
  component describes.}
\label{f:mg_model}
\end{center}
\end{figure}

%magnesium

\citet{1998ApJ...496..908W} identified the strong and relatively
isolated absorption feature at 1.05 \um\ as \mgii\ \lam1.0927 \um.
The line is expected to be observed with decreasing velocity.  The
velocity is then expected to cease changing when the photosphere has
receded below the inner edge of the \mgii\ distribution.  Magnesium is
a product of explosive carbon burning, but not oxygen burning, as the
required higher temperatures for oxygen burning generate products
further along the alpha chain, such as, silicon and sulfur.  This
makes the \mgii\ line a sensitive probe to the location of the inner
edge of carbon burning in velocity space \citep{1998ApJ...496..908W,
  2001RMxAC..10..190M, 2002ApJ...568..791H, 2009AJ....138..727M}.
Here, we focus on measuring the velocity using the location of the
absorption minimum of the \mgii\ \lam1.0927 \um\ line profile.  With
the NIR spectra of SN~2011fe and previous data from the literature, we
explore the range of \mgii\ velocity and the accuracy limit of the
velocity measurements.

%inadequacy of Gaussian line profile

The canonical Gaussian line profile is inadequate in fitting the
\mgii\ P-Cygni profile, which contains influences from the P-Cygni
wings of \ion{C}{1}, \ion{O}{1} and \ion{Si}{2}.  Using the Gaussian
function to fit supernova line profiles often results in poor model
fits which are unaccounted for by the errors.  To properly estimate
the uncertainties in the velocity measurements, we opted to use a
principal component model which was built from data and empirically
describes the variation in the line profiles. The formulation and the
construction of the model are described in the appendix.

%model spectra

The first four principal components and the model spectra, which
account for over 90\% of the variations, are plotted in
Figure~\ref{f:mg_model}.  The first two principal components, $\xi_0$
and $\xi_1$, show strong correlations between the strengths of the
\mgii\ \lam\lam1.0092, 1.0927 \um\ lines and the \ci\ \lam1.0693
\um\ line. These line strengths are also correlated with those of
\ion{O}{1} and \ion{Si}{3} lines further to the red.  The next
principal component, $\xi_2$, describes approximately the varying
tilts in the input spectra.  Using these principal components, the
model spectrum of the \mgii\ feature was then built as
$M(\boldsymbol{p},n)$, which varies as a function of the number of
principal components $n$ and the projections $\boldsymbol{p}$ onto the
corresponding principal components.

%fitting technique

To fit the observed spectrum, denoted by $O(\lambda)$ and with
uncertainty $\Delta O(\lambda)$ and $m$ wavelength elements, we
allowed extra freedom in the model spectra $M(n,\boldsymbol{p})$ in
the normalization $C$ and wavelength shift $\lambda_{\rm{shift}}$, and
minimized the following:
\begin{equation}
\chi^2 = \sum_{i=0}^{m-1} \left[ 
\frac{O_i(\lambda)-M(n,\boldsymbol{p},C,\frac{\lambda_{\rm{shift}}}{1.05\rm{\mu m}}\lambda)}
{\Delta O_i(\lambda)} \right]^2.
\label{e:chisq}
\end{equation}
The best fit model spectrum was determined using the IDL package
\texttt{mpfit} \citep{2009ASPC..411..251M}, a nonlinear least-squares
fitting program ported from the FORTRAN package \texttt{MINPACK-1}.
The number of principal components to include, $n$, was fixed in the
fit and chosen, such that, the reduced $\chi^2$ of the resulting best
fit is closest to 1, typically $n=6-8$.

%determination of minima

Once the best-fit model spectrum was found, we attempted to remove the
steep ``continuum'' with methods similar to that employed by
\citet{2009AJ....138..727M}.  A straight line connecting the model
spectrum at fixed wavelengths to the blue and red side of best-fit
$\lambda_{\rm{shift}}$ was taken as the continuum and removed from the
model spectrum.  The profile minimum of the continuum-removed model
spectrum was then used to determine the line velocity of \mgii.  The fit
uncertainty of $\lambda_{\rm{shift}}$ was used to derive the
measurement error for the velocity.

%Mg II velocity vs time

\begin{figure}
\begin{center}
\epsscale{1.2}
\plotone{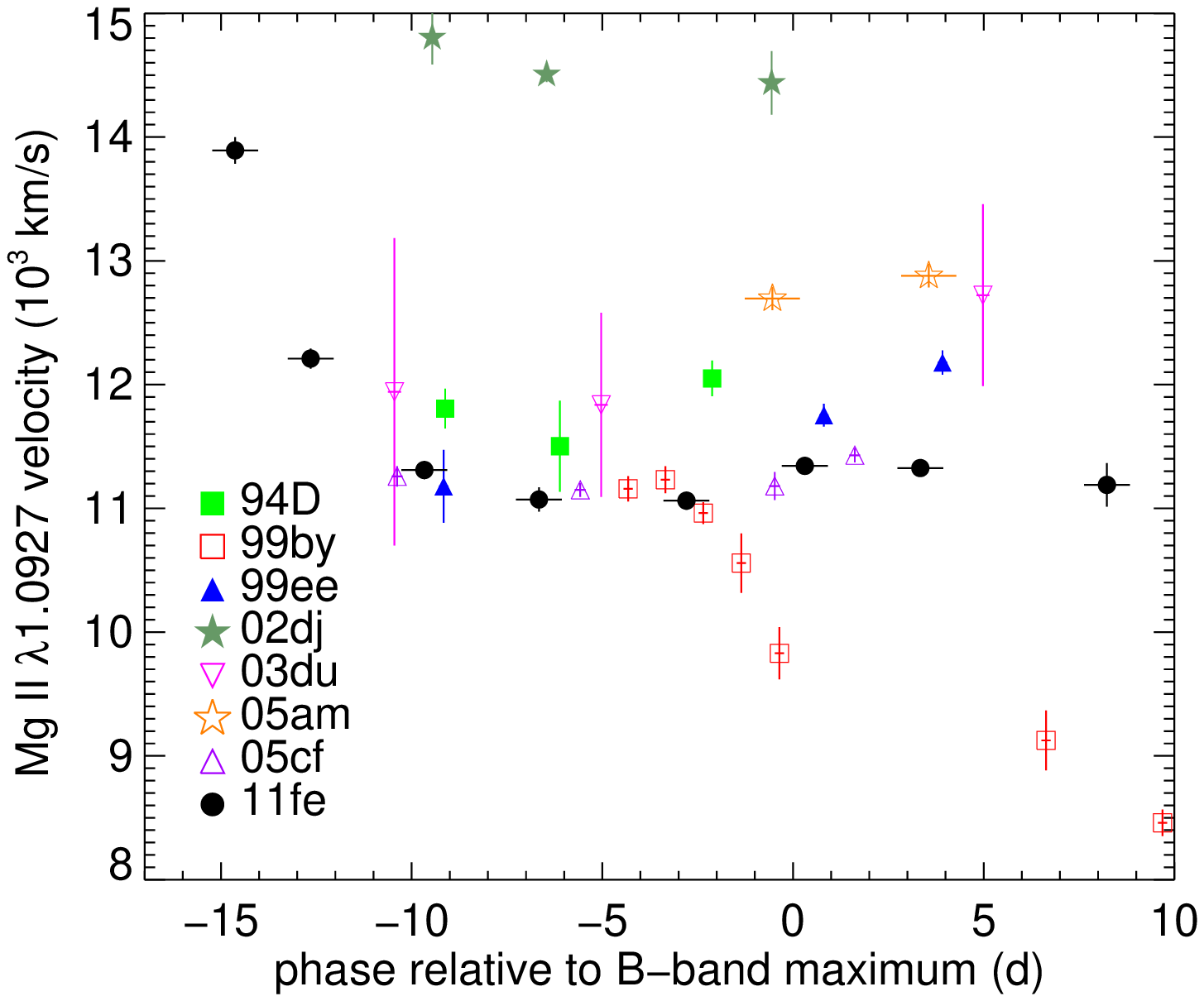}
\caption{The time evolution of the \mgii\ \lam1.0927 \um\ velocity.}
\label{f:mg_t}
\end{center}
\end{figure}

The \mgii\ velocities from several literature early-phase time-series
NIR spectra were determined and plotted with the velocities of
SN~2011fe in Figure~\ref{f:mg_t}.  As predicted by
\citet{1998ApJ...496..908W}, SN~2011fe shows a rapid decrease in the
\mgii\ line velocity followed by an extended epoch of constant
velocity, beginning at $\sim10$ days before maximum and lasting until
the feature disappears at $\sim10$ days past maximum.  The velocity
during this phase is remarkably constant for SN~2011fe, showing a
dispersion of only 130 \kms, on the order of the velocity resolution
of these medium-resolution spectra.  The emission portion of the
\ci\ \lam1.0693 \um\ P-Cygni profile is blended with the absorption
feature of \mgii\ \lam1.0927 \um\ (Section~\ref{s:carbon}).  In the
case of SN~2011fe, the time-varying characteristic of the \ci\ feature
had very little influence on the \mgii\ velocity during this epoch.
For the rest of the SNe~Ia in our sample, they appear to be caught
after the \mgii\ line has entered the epoch of constant velocity, with
the exception of the subluminous SN~1991bg-like SN~1999by. For
SN~1999by, the \mgii\ velocity continues to decrease well past maximum
light and indicates a distribution of \mgii\ down to $\sim8,000$ \kms.
Note that SN~2002dj is a significant outlier in Figure~\ref{f:mg_t}
with exceptionally high \mgii\ velocity which persisted for a wide
phase range before maximum light.  SN~2002dj is also identified as a
high velocity gradient object in the \citet{2005ApJ...623.1011B}
classification \citep{2008MNRAS.388..971P}, similar to SN~2002bo
\citep{2004MNRAS.348..261B}.  The NIR spectrum of SN~2002bo also shows
high \mgii\ velocity, although not as extreme as that of SN~2002dj.
These results may point to a correspondence between high \ion{Si}{2}
velocity gradients and high \mgii\ velocities.

\begin{figure}
\begin{center}
\epsscale{1.2}
\plotone{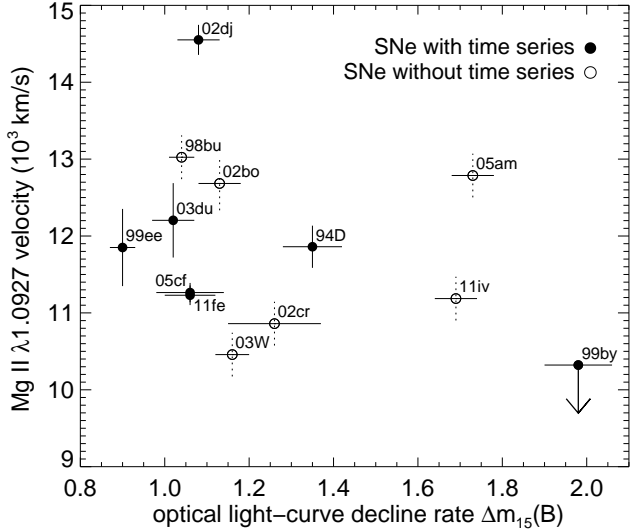}
\caption{The \mgii\ \lam1.0927 \um\ velocity versus optical
  light-curve decline rate \dm.  SNe~Ia with three or more NIR spectra
  taken during the constant velocity epoch are plotted in filled
  circles, while the rest are plotted in open circles.  For the
  peculiar SN~1999by, a downward arrow indicates that the constant
  velocity epoch has yet to be reached in the defined time interval
  (see Figure~\ref{f:mg_t}).}
\label{f:mg_dm15}
\end{center}
\end{figure}

%Mg II velocity vs dm15

Here, we test whether the location of the inner edge of explosive
carbon burning has a significant impact on the $^{56}$Ni productions
of SNe~Ia.  Using \dm\ \citep{1993ApJ...413L.105P} as a proxy for the
amount of $^{56}$Ni produced, the result is plotted in
Figure~\ref{f:mg_dm15}.  The \mgii\ velocity for each SN is the mean
velocity, weighted by the measurement errors, in its constant epoch.
The uncertainty is the intrinsic dispersion of the velocity around the
mean during the constant epoch.  For those SNe without adequate
time-series data, the uncertainty is assumed to be the mean of the
dispersions of all SNe~Ia with time series, excluding SN~1999by.

%poor correlation with dm15

There is a large spread in the locations of the inner edge of the
\mgii\ distribution in SNe~Ia.  For a range of normal SNe~Ia
$1.0<$\dm$<1.2$, the spread of \mgii\ velocities is as high as 5,000
\kms, as was observed by \citet{2003ApJ...591..316M,
  2009AJ....138..727M}.  There appears to be no correlation between
the \mgii\ velocity and \dm.  This result is surprising, since the
transition density is expected to have a strong influence on the yield
of $^{56}$Ni \citep[e.g.,][]{1995ApJ...444..831H}.  The model of
\citet{2002ApJ...568..791H} for SN~1999by corresponds to the lowest
transition density in the model grid.  SN~1999by also happens to be a
subluminous event and has very low \mgii\ velocities.  Yet SN~2005am,
with the next fastest decline rate in our sample, has one of the
highest \mgii\ velocities measured.  If the \mgii\ velocity is indeed
a strong function of the transition density, its effect on the
$^{56}$Ni production and the brightness of SNe~Ia appears to be
secondary.  Since its correlation to \dm\ is weak, it provides
independent information which may affect the brightness of SNe~Ia.  We
explore the possibility of using the \mgii\ velocity measurement to
further improve the use of SNe~Ia as distance indicators in
Section~\ref{s:conc}.

%%%%%%%%%%%%%%%%
%% Iron group %%
%%%%%%%%%%%%%%%%

\subsection{H-band Iron-peak Feature}
\label{s:iron}

%exposed core

Soon past $B$-band maximum light, the NIR spectra take on a
characteristic shape.  Deep depression in regions between $1.2 - 1.5$
\um\ is coupled with a dramatic feature between $1.5 - 1.9$ \um,
seemingly in emission (Figure~\ref{f:spec}).  The feature was first
noted by \citet{1973ApJ...180L..97K}.  We now understand the deep
depression as the result of a lack of line-blanketing opacity
\citep{1994MNRAS.266L..17S}.  The early NIR spectra of SNe~Ia are
dominated by electron scattering.  As the ejecta expand and cool, the
electron scattering opacity decreases and, in turn, lowers the
continuum flux level and exposes the core.  In contrast, the strong
line blanketing opacity from iron-peak elements causes the prominent
feature at $1.5 - 1.9$ \um\ to form at a larger effective radius and
appear in emission \citep{1998ApJ...496..908W}.  The strongly-variable
opacity provides probes of very different depths at the same epoch.
An alternate origin of the increased flux at $1.5-1.9$ \um\ was
proposed to be increased fluorescence in the region by
\citet{2012MNRAS.427..994G}.

%in SN 2011fe

In the time-series data of SN~2011fe, we caught the budding of the
prominent iron-peak feature in the spectrum at $\sim3$ days past
$B$-band maximum (Figure~\ref{f:spec}).  It then emerges rapidly to be
the dominant feature in the NIR.  This feature coincides in wavelength
with the $H$ band.  The rapid time evolution and the shear size of the
feature inevitably cause large uncertainties in the $H$-band
k-corrections.  Understanding the evolution and the diversity of the
feature is crucial to control the systematic errors originating from
k-correction estimates.  The relative flux levels in regions of
contrasting opacity also provide a temperature diagnostic for the
line-blanketing material.  Here, we explore the flux ratio across the
break at 1.5 \um, which separates the different opacity regimes.

\begin{figure}
\begin{center}
\epsscale{1.2}
\plotone{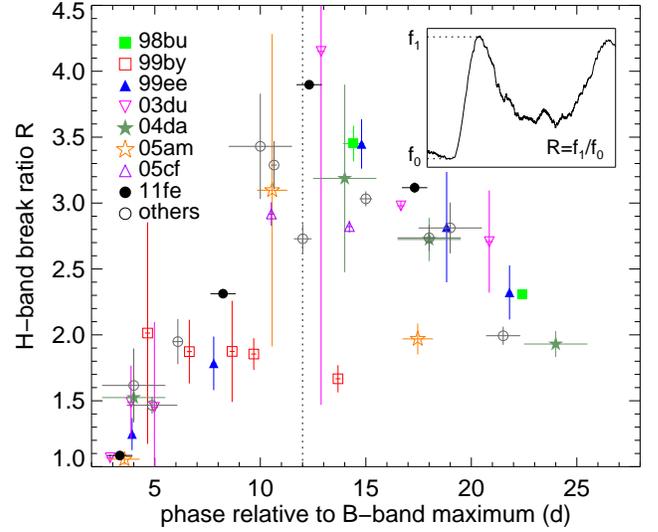}
\caption{Time evolution of the ratio across the break at 1.5 \um\
  $R$. The inset in the top right corner illustrates the definition of
  the break ratio with the spectrum SN~2011fe at 12.3 days past
  maximum.}
\label{f:hbreak_t}
\end{center}
\end{figure}

\begin{figure}
\begin{center}
\epsscale{1.2}
\plotone{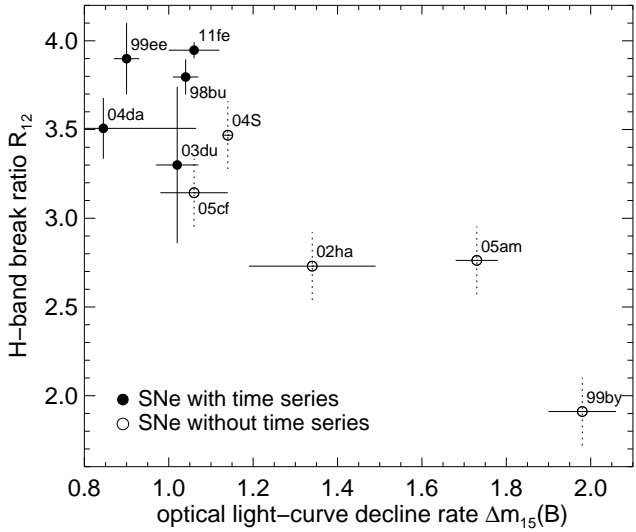}
\caption{The ratio across the break at 1.5 \um\ at 12 days past
  maximum, $R_{12}$, versus optical light-curve decline rate
  \dm. SNe~Ia with two or more time-series observations in the
  post-peak decline are plotted with filled circles, while the rest
  are plotted with open circles.}
\label{f:hbreak_dm15}
\end{center}
\end{figure}

%H-band break ratio versus time

We define the $H$-band break ratio $R$ as the ratio of the maximum
flux just redward of the 1.5 \um\ break to the minimum flux just
blueward of the 1.5 \um\ break (Figure~\ref{f:hbreak_t}).  To
determine the flux ratios for the sample of NIR spectra, the spectra
were first smoothed with a Gaussian filter with a fixed width of
$\Delta\lambda/\lambda=0.0025$ in logarithmic wavelength space.  The
minima and maxima were found in fixed wavelength regions from the
smoothed spectra.  Since the velocity of the feature and the
associated uncertainty do not enter in this analysis, we found that
the Gaussian smoothing technique to be simpler and more robust than
the PCA models described in Section~\ref{s:magnesium}.  The
measurement errors were estimated from the dispersions in the
neighboring pixels, within $\pm100$\AA, of the minima and maxima.  The
signal-to-noise ratio of each spectrum around the H-band break region
is therefore reflected in the error.  The resulting $H$-band break
ratios for SN~2011fe and the sample of literature spectra are plotted
in Figure~\ref{f:hbreak_t}.  The $H$-band features of SNe~Ia appear to
emerge at very similar phases, $\sim 3$ days past $B$-band maximum
light.  With the exception of the peculiar SN~1999by, the flux ratios
appear to peak uniformly at $\sim 12$ days past $B$-band maximum
light; although, more time-series data are required to characterize
the rise.  Then, a remarkably uniform linear decline is observed among
normal SNe~Ia.  Excluding the peculiar SN~1999by, we determined an
error-weighted mean linear decline of $-0.146$ day$^{-1}$ with a
standard deviation of $0.036$ day$^{-1}$.

%H-band break ratio versus dm15

From our limited sample, the time evolution of the $H$-band break
ratio $R$ appears to be uniform with a range of values at peak.  We
defined $R_{12}$ as the $H$-band break ratio at 12 days past $B$-band
maximum.  For each SN~Ia with time-series data after the ratio peak, a
straight line was fit to the linear post-peak decline.  Then $R_{12}$
was determined by extrapolating on the best-fit line to 12 days past
$B$-band maximum.  The uncertainty in $R_{12}$ was derived from the
error in the slope of the linear fit.  For SNe~Ia without adequate
time-series data to determine the linear decline of the $H$-band break
ratio, the error-weighted mean decline rate of $-0.146$ day$^{-1}$ and
an average uncertainty were adopted.  The resulting $R_{12}$
measurements are plotted against \dm\ in Figure~\ref{f:hbreak_dm15}.

%strong correlation between H-band break and dm15

The peak $H$-band Break ratio shows strong correlation with
\dm\ (Figure~\ref{f:hbreak_dm15}); although, more time-series data are
needed to confirm this finding.  If the effect from intermediate-mass
material at $1.2 - 1.5$ \um\ is secondary, the correlation describes a
temperature sequence.  The temperature of the iron-peak material
appears to be closely tied to the production of $^{56}$Ni, more so
than the transition density (Section~\ref{s:magnesium}).  In the view
point of contrasting opacity in the region, the observed effect can be
explained as follows.  At $\sim12$ days past $B$-band maximum, the
electron scattering opacity is low enough that the iron-peak feature
is fully exposed.  The temperature of the iron-peak material now
dictates the size of this feature.  As the ejecta expands, the
line-blanketing opacity drops with the decreasing temperature of the
iron-peak material.  The iron-peak feature at $1.5 - 1.9$ \um\ then
forms at decreasing effective radii and appears weaker.  The
uniformity of the time evolution indicates that the temperature
decline of the iron-peak material is consistent in normal SNe~Ia,
regardless of the varying amounts of $^{56}$Ni produced.  The
predictability of the strength and the evolution of the iron-peak
feature means that we can indeed improve upon the current method of
NIR k-correction \citep{2007ApJ...663.1187H, 2009PhDT.......228H}, by
building, for example, spectral templates as a function of \dm.

%%%%%%%%%%%%%%%%%%%
%% K-corrections %%
%%%%%%%%%%%%%%%%%%%

\section{NIR K-corrections}
\label{s:kcorr}

\begin{figure}
\begin{center}
\epsscale{1.2}
\plotone{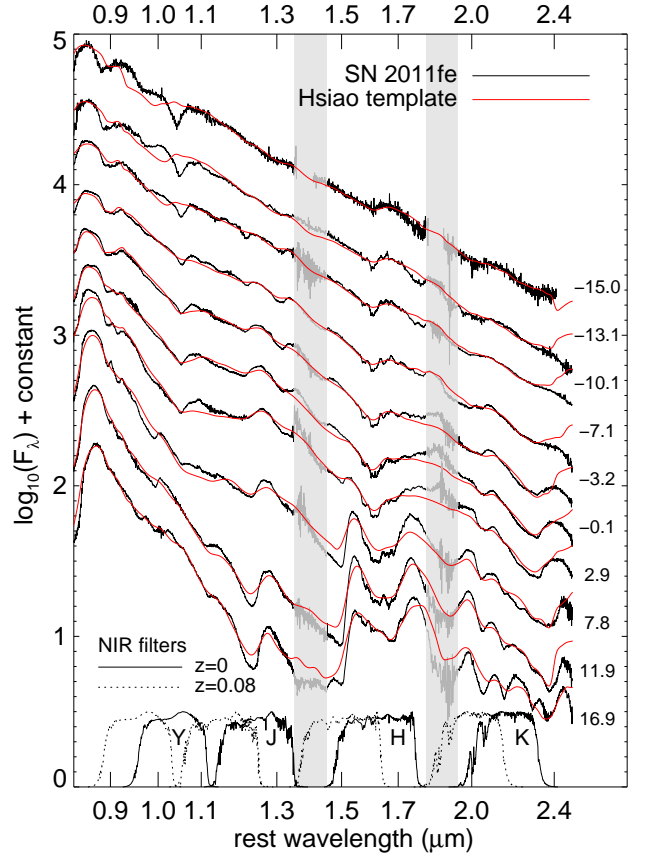}
\caption{Comparison between the NIR spectra of SN~2011fe and the
  revised template of \citet{2007ApJ...663.1187H} warped to match
  observed colors from \izyjhk\ photometry. At the bottom, the
  locations of the observed NIR filters at $z=0$ and $z=0.08$ are
  plotted in the rest frame of the supernova.}
\label{f:kcorr_spec}
\end{center}
\end{figure}

\begin{figure}
\begin{center}
\epsscale{1.2}
\plotone{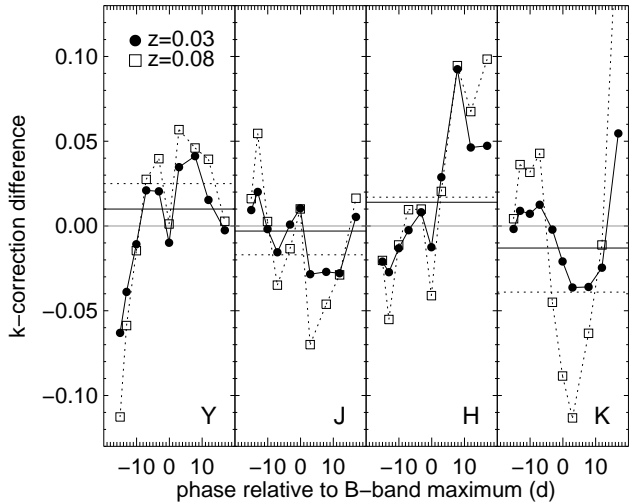}
\caption{NIR \yjhk\ k-correction difference between SN~2011fe and the
  revised template of \citet{2007ApJ...663.1187H}. Two redshifts are
  shown: $z=0.03$ (filled circles and solid lines) and $z=0.08$ (open
  squares and dotted lines). The horizontal lines mark the estimated
  offsets in the peak magnitude caused by the k-correction difference
  for $z=0.03$ (solid lines) and $z=0.08$ (dotted lines).}
\label{f:kcorr_diff}
\end{center}
\end{figure}

%k-corrections

K-corrections account for the effect of cosmological expansion.
SNe~Ia at different redshifts would have their spectral energy
distribution (SED) sampled at different wavelength regions by the same
observed filter.  A k-correction converts an observed magnitude to
that which would be observed in the rest frame of the same or another
filter, allowing for the comparison of the brightness of SNe~Ia at
various redshifts \citep{1968ApJ...154...21O}\footnote{See
  \citet{2002SPIE.4836...61A} for an alternative experimental approach
  which eliminates the need for k-corrections.}.  The calculation of
k-corrections requires knowledge of the SED.  While the effect of an
assumed spectral template has been extensively studied in the optical
\citep{2002PASP..114..803N, 2007ApJ...663.1187H}, the effect is not
well understood in the NIR due to the relatively small spectroscopic
sample \citep{2009PhDT.......228H}.

%comparing spectra and template

The current method of NIR k-corrections employs the revised spectral
templates of \citet{2007ApJ...663.1187H}, warped to match the observed
colors of the SN~Ia \citep{2011AJ....141...19B}.  Here, we explore the
effect of assuming the spectral template for the case of a
SN~2011fe-like object at various redshifts.  Before the k-corrections
were calculated, the spectral template at each phase was warped to the
NIR colors provided by WFCAM and WHIRC \izyjhk\ photometry of
SN~2011fe \citep[Im et al. in prep;][]{2012ApJ...754...19M}.  The NIR
spectra of SN~2011fe and the color-corrected spectral templates are
plotted together for comparison in Figure~\ref{f:kcorr_spec}.

%spectral feature mismatches

The mismatch of the \mgii\ feature near 1.05 \um\ is evident in the
first three spectra.  The discrepancy is not surprising, since these
are the earliest NIR spectra published to date.  There also appears to
be some mismatch in the strength of the prominent $H$-band feature at
later phases.  This illustrates the limitation of using a single
template for all SNe~Ia.  As shown in Section~\ref{s:iron}, the
strength of this iron-peak feature correlates with \dm.  Armed with
this well-defined correlation, one can build a sequence of templates
as a function of \dm\ to minimize the k-correction uncertainty
associated with this feature.

%k-correction differences

The mismatches between the spectra and template are reflected in the
k-correction differences plotted in Figure~\ref{f:kcorr_diff}.  The
k-correction differences were calculated at two redshifts, 0.03 and
0.08, which coincide with the redshift limits of current surveys of
nearby SNe~Ia in the Hubble flow.  The larger differences observed at
early phases in the $Y$ band and at late phases in the $H$ band
originate from the mismatches in the \mgii\ feature at early phases
and the iron-peak feature at late phases.  At $z=0.08$ where the rest
and observed filters are significantly misaligned, the mean absolute
sizes of the differences over the phase range are 0.04, 0.03, 0.04,
0.06 mag in $Y$, $J$, $H$, $K$, respectively, similar to the numbers
reported by \citet{2007ApJ...663.1187H} for the optical filters.  The
offsets in the peak magnitude caused by assuming the spectral template
can be approximated by using the average of the k-correction
differences weighted by the light-curve template errors.  For a
SN~2011fe-like object, the impact of assuming a template SED on the
determination of peak magnitude is small, on the level of 0.01 mag for
$z=0.03$ and $0.02-0.04$ for $z=0.08$ (Figure~\ref{f:kcorr_diff}).

%correlation in time

Nonetheless, whether the differences at various phases correlate with
each other offers us important clues as to how to proceed to minimize
the impact of uncertainties in k-corrections.  If the k-correction
differences from various phases are completely uncorrelated, more
observations on the light curve could drive down the uncertainty in
the peak magnitude determination.  If the differences are
significantly correlated, there is room for improvement in the time
evolution of the spectral templates.  The first attempt to quantify
these correlations in the optical was made by
\citet{2007ApJ...663.1187H}.  This study requires a large sample of
time-series spectra, which does not yet exist in the NIR.  In
Section~\ref{s:conc}, we describe the ongoing efforts to build such a
data set.

%warping the template

The availability of color constraints from neighboring filters is also
crucial for obtaining good k-correction estimates.  For example, an
accurate k-correction to the $J$ band requires $Y-J$ and $J-H$ colors
to warp the spectral template to the correct observed colors around
the wavelength region in question.  Missing color information inflates
the k-correction uncertainty to $\sim0.1$ mag
\citep{2007ApJ...663.1187H}.  The higher k-correction difference seen
in the $K$ band in Figure~\ref{f:kcorr_diff} is most likely due to the
lack of constraining observation in a redder band.

%%%%%%%%%%%%%%%%%
%% Conclusions %%
%%%%%%%%%%%%%%%%%

\section{Conclusions}
\label{s:conc}

%outline of major findings

We have presented ten medium-resolution, high signal-to-noise ratio
NIR spectra of SN~2011fe.  This data set constitutes the earliest
time-series NIR spectroscopy of a SN Ia, with the first spectrum
obtained at only 2.58 days past the explosion.  We take advantage of
these densely-sampled observations to gain insights into the time
evolution and diversity of SN Ia NIR spectral features.  The main
findings are outlined below:

\begin{enumerate}
  \item With the aid of the automated spectrum synthesis code
    \texttt{SYNAPPS}, \ci\ \lam1.0693 \um\ is detected with increasing
    strength up to maximum light in SN~2011fe. The delay in its onset
    demonstrates the potential of the NIR \ci\ line to be a superior
    tracer of unprocessed material to the more commonly used optical
    \cii\ \lam0.6580 \um\ line.
  \item The \mgii\ velocity of SN~2011fe, measured from the absorption
    minimum of the \mgii\ \lam1.0927 \um\ line profile, decreases
    rapidly at very early phases, then flattens to a remarkably
    constant evolution from 10 days before to 10 days past $B$-band
    maximum, with a dispersion of only 130 \kms.
  \item Taking advantage of the constant velocity over a wide time
    window, the \mgii\ \lam1.0927 \um\ line, even without time-series
    observations, can be used to locate the inner edge of carbon
    burning.  The \mgii\ profiles of SNe~Ia exhibit a wide spread of
    velocities, as high as 5,000 \kms, and show no correlation with
    \dm.
  \item The prominent feature at $1.5-1.9$ \um\ shows uniform phase
    evolution in SNe~Ia.  The flux ratio across the break at 1.5
    \um\ increases from $\sim3$ days past maximum, when the break
    appears, and reaches its peak at $\sim12$ days past maximum.  As a
    proxy for the temperature of the line-blanketing material, the
    strong correlation between the peak flux ratio and \dm\ indicates
    a close tie between temperature of the iron-peak elements and the
    production of $^{56}$Ni.
  \item Despite the presence of the prominent and rapidly-evolving
    feature at $1.5-1.9$ \um\ near maximum light, the predictability
    of its strength and the evolution means that NIR k-corrections can
    be improved by producing spectral templates as a function of \dm.
\end{enumerate}

%connection of Mg II velocity to metallicity

The result that the \mgii\ velocity does not correlate with \dm\ is an
intriguing one, since the transition density is expected to have a
strong influence on the yield of $^{56}$Ni
\citep[e.g.,][]{1995ApJ...444..831H}.  Is it possible that the
\mgii\ velocity describes variations in SN~Ia luminosity on a
secondary level?  Recently, SN~Ia luminosity has been found to have a
dependence on the host-galaxy stellar mass, in addition to the
well-known dependence on light-curve shape and colors
\citep[e.g.,][]{2010ApJ...715..743K, 2010MNRAS.406..782S,
  2010ApJ...722..566L}. It has been postulated that the effect may be
attributed to the impact of progenitor metallicity on $^{56}$Ni
production \citep[e.g.,][]{2003ApJ...590L..83T, 2010ApJ...720...99J}.
With the transition density linked to progenitor metallicity
\citep{2010ApJ...720...99J}, the NIR \mgii\ velocity is potentially a
more direct probe of progenitor metallicity and a better calibrator of
SN~Ia luminosity than host-galaxy stellar mass.

%study of MgII lines with hubble flow objects, snapshot program

With the limited size of the current NIR spectroscopic sample, the
\mgii\ velocity shows a weak trend with the host-galaxy stellar mass,
hinting a link of \mgii\ velocity to progenitor properties; although,
a larger sample with unbiased host-galaxy properties is needed to
determine if such a correlation indeed exists.  To test whether the
\mgii\ velocity provides better calibration of SN~Ia luminosity than
host-galaxy stellar mass, a statistical sample of NIR spectra,
especially those of SNe~Ia in the Hubble flow, is required.  With
improved NIR spectrographs available on large telescopes, obtaining
NIR spectra of SNe~Ia in the Hubble flow is now feasible.  For
example, a pre-maximum NIR spectrum of SN~2012ar, a SN~Ia with a
recession velocity of 8472 \kms, was obtained in the low-resolution
mode with the Folded-port Infrared Echellette (FIRE) on the 6.5-meter
Magellan Telescope.  The spectrum reaches a signal-to-noise ratio of
$>50$ at 1 \um, with an on-target integration time of only 10 minutes.
With this signal-to-noise ratio, the measurement error for the
\mgii\ velocity is still well below the intrinsic velocity dispersion
of SN~2011fe in the constant velocity epoch
(Section~\ref{s:magnesium}).

%growing sample of NIR time-series spectroscopy

Throughout the paper, we have emphasized the need for an improved data
set of NIR spectra to study this relatively unexplored area and to
confirm the various findings from this paper.  Starting in late 2011,
we began a five-year program to build a large sample of high quality
SN Ia NIR spectra using NIR spectrographs on large telescopes, such as
FIRE, GNIRS and ISAAC (Phillips et al. in prep).  In the first year of
the program, 130 NIR spectra have been obtained of 23 SNe~Ia.  The NIR
time-series spectroscopy of SN~2011fe has demonstrated the many
utilities of NIR spectra, both before and after maximum light, for
both time-series and single ``snapshot'' observations.  This improved
data set will allow us to examine the efficacy of these tools and help
to further decipher the nature of SNe~Ia in the near future.

%%%%%%%%%%%%%%%%%%%%%
%% Acknowledgments %%
%%%%%%%%%%%%%%%%%%%%%

\begin{acknowledgments} 

\bigskip

This paper is based on data obtained with the Gemini Observatory,
under the long-term program GN-2011B-Q-68, the NASA Infrared Telescope
Facility (IRTF), the United Kingdom Infrared Telescope (UKIRT), and
the 6.5-meter Magellan Telescopes.  UKIRT is operated by the Joint
Astronomy Centre on behalf of the Science and Technology Facilities
Council of the UK.  We would like to thank Nancy Levenson for
approving our GNIRS observations to be performed outside of the
proposed observing period.  Most observations were obtained at
facilities on Mauna Kea.  The authors would like to recognize the very
significant cultural role and reverence that the summit of Mauna Kea
has within the indigenous community of Hawaii.  We are grateful for
our opportunity to conduct observations from this mountain.

This material is based upon work supported by the National Science
Foundation under Grant No. AST-1008343.  M.~S. acknowledges the
generous support provided by the Danish Agency for Science and
Technology and Innovation through a Sapere Aude Level 2 grant.
M.~I. and Y.~J. acknowledge the support from the Creative Initiative
program, No. 2010-0000712, of the National Research Foundation of
Korea (NRFK).  G.~P. acknowledges support from the Millennium Center
for Supernova Science through grant P10-064-F funded by ``Programa
Bicentenario de Ciencia y Tecnolog\'ia de CONICYT'' and ``Programa
Iniciativa Cient\'ifica Milenio de MIDEPLAN''.

This research used resources from the National Energy Research
Scientific Computing Center (NERSC), which is supported by the Office
of Science of the U.S. Department of Energy under Contract
No. DE-AC02-05CH11231.  We have also made use of the NASA/IPAC
Extragalactic Database (NED) which is operated by the Jet Propulsion
Laboratory, California Institute of Technology, under contract with
the National Aeronautics and Space Administration.

\end{acknowledgments}

%%%%%%%%%%%%%%%%%%
%% Bibliography %%
%%%%%%%%%%%%%%%%%%

%%%%%%%%%%%%%%
%% Appendix %%
%%%%%%%%%%%%%%

\appendix
\section{Principal component model line profile}
\label{s:pca}

%principal component analysis of data

Principal component analysis (PCA) is a statistical technique useful
for reducing the dimensionality and identifying patterns in a set of
multi-dimensional data. We use the ``bra ket'' notation of
\citet{2006ApJS..163..110S} here for the formulation of PCA.  A sample
of early, high signal-to-noise ratio NIR spectra from the literature,
including those of SN~2011fe, were used as PCA data input to build the
model of the \mgii\ \lam1.0927 \um\ profile.  Each spectrum was
shifted in wavelength, such that the \mgii\ line minima coincide at a
fiducial wavelength of 1.05 \um, normalized to the same integrated
flux in the $Y$ band, and interpolated into $N$ fixed wavelength
elements.  The wavelength interval was chosen to be 10\AA,
approximately matching and preserving the medium resolution of the
SN~2011fe spectra at 1.05 \um.  The log of the flux was used as the
PCA input data and is denoted by $|m \rangle$.  The mean spectrum
$|\mu \rangle$ was computed and subtracted from the data set.  A $N
\times N$ covariance matrix was then computed to determine the
strength of the correlation between spectral features.

%eigenvectors and eigenvalues

The eigenvectors and eigenvalues of the covariance matrix were
computed using internal IDL routines which utilize Householder
reductions and the QL method \citep{1992nrca.book.....P}.  Each of the
$N$ eigenvectors, or principal components, denoted here by $|\xi_j
\rangle$, was ranked by the corresponding eigenvalue, which provides a
measure of the amount of data variation the principal component
describes.  The zeroth-ranked principal component points in the
direction of maximum variance in the $N$-dimensional data space, and
subsequent principal components are orthogonal vectors:
\begin{equation}
\langle \xi_i | \xi_j \rangle = \delta_{ij}.
\end{equation}
Each of our input spectra can be expressed as the sum of the
eigenvectors:
\begin{equation}
|m_i \rangle = |\mu \rangle + \sum_{j=0}^{N-1} p_{ij}|\xi_j \rangle,
\end{equation}
where $p_{ij}$ is the projection of the $i$th input spectrum on the
$j$th eigenvector.

%model spectrum

Using these eigenvectors, the model spectrum of the \mgii\ feature was
then built as $M(\boldsymbol{p},n)$, which varies as a function of the
number of eigenvectors $n \leq N$, and projection $p_j$ for each
eigenvector $|\xi_j \rangle$:
\begin{equation}
M(n,\boldsymbol{p}) = |\mu \rangle + \sum_{j=0}^{n-1} p_j|\xi_j \rangle.
\label{e:model}
\end{equation}


\begin{thebibliography}{}
\bibitem[Aldering et al.(2002)]{2002SPIE.4836...61A} Aldering, G., Adam, 
G., et al.\ 2002, \procspie, 4836, 61 
\bibitem[Aspden et al.(2008)]{2008ApJ...689.1173A} Aspden, A.~J., Bell, 
J.~B., Day, M.~S., Woosley, S.~E., \& Zingale, M.\ 2008, \apj, 689, 1173 
\bibitem[Barone-Nugent et al.(2012)]{2012MNRAS.425.1007B} Barone-Nugent, 
R.~L., Lidman, C., et al.\ 2012, \mnras, 425, 1007 
\bibitem[Benetti et al.(2004)]{2004MNRAS.348..261B} Benetti, S., Meikle, 
P., et al.\ 2004, \mnras, 348, 261 
\bibitem[Benetti et al.(2005)]{2005ApJ...623.1011B} Benetti, S., 
Cappellaro, E., et al.\ 2005, \apj, 623, 1011 
\bibitem[Blondin et al.(2012)]{2012AJ....143..126B} Blondin, S., Matheson, 
T., et al.\ 2012, \aj, 143, 126 
\bibitem[Bloom et al.(2012)]{2012ApJ...744L..17B} Bloom, J.~S., Kasen, D., 
et al.\ 2012, \apjl, 744, L17 
\bibitem[Bowers et al.(1997)]{1997MNRAS.290..663B} Bowers, E.~J.~C., 
Meikle, W.~P.~S., et al.\ 1997, \mnras, 290, 663 
\bibitem[Branch et al.(2005)]{2005PASP..117..545B} Branch, D., Baron, E., 
Hall, N., Melakayil, M., \& Parrent, J.\ 2005, \pasp, 117, 545 
\bibitem[Branch et al.(2003)]{2003AJ....126.1489B} Branch, D., Garnavich, 
P., et al.\ 2003, \aj, 126, 1489 
\bibitem[Branch et al.(2007)]{2007PASP..119..709B} Branch, D., Troxel, 
M.~A., et al.\ 2007, \pasp, 119, 709 
\bibitem[Brown et al.(2012)]{2012ApJ...753...22B} Brown, P.~J., Dawson, 
K.~S., et al.\ 2012, \apj, 753, 22 
\bibitem[Burns et al.(2011)]{2011AJ....141...19B} Burns, C.~R., 
Stritzinger, M., et al.\ 2011, \aj, 141, 19 
\bibitem[Chomiuk et al.(2012)]{2012ApJ...750..164C} Chomiuk, L., Soderberg, 
A.~M., et al.\ 2012, \apj, 750, 164 
\bibitem[Cushing et al.(2004)]{2004PASP..116..362C} Cushing, M.~C., Vacca, 
W.~D., \& Rayner, J.~T.\ 2004, \pasp, 116, 362 
\bibitem[Elias et al.(1985)]{1985ApJ...296..379E} Elias, J.~H., Matthews, 
K., Neugebauer, G., \& Persson, S.~E.\ 1985, \apj, 296, 379 
\bibitem[Elias et al.(1998)]{1998SPIE.3354..555E} Elias, J.~H., 
Vukobratovich, D., et al.\ 1998, \procspie, 3354, 555 
\bibitem[Filippenko(1982)]{1982PASP...94..715F} Filippenko, A.~V.\ 1982, 
\pasp, 94, 715 
\bibitem[Folatelli et al.(2010)]{2010AJ....139..120F} Folatelli, G., 
Phillips, M.~M., et al.\ 2010, \aj, 139, 120 
\bibitem[Folatelli et al.(2012)]{2012ApJ...745...74F} Folatelli, G., 
Phillips, M.~M., et al.\ 2012, \apj, 745, 74 
\bibitem[Foley et al.(2012)]{2012ApJ...753L...5F} Foley, R.~J., Kromer, M., 
et al.\ 2012, \apjl, 753, L5 
\bibitem[Frogel et al.(1987)]{1987ApJ...315L.129F} Frogel, J.~A., Gregory, 
B., et al.\ 1987, \apjl, 315, L129 
\bibitem[Gall et al.(2012)]{2012MNRAS.427..994G} Gall, E.~E.~E., 
Taubenberger, S., et al.\ 2012, \mnras, 427, 994 
\bibitem[Gamezo et al.(2004)]{2004PhRvL..92u1102G} Gamezo, V.~N., Khokhlov, 
A.~M., \& Oran, E.~S.\ 2004, Physical Review Letters, 92, 211102 
\bibitem[Gamezo et al.(2003)]{2003Sci...299...77G} Gamezo, V.~N., Khokhlov, 
A.~M., Oran, E.~S., Chtchelkanova, A.~Y., 
\& Rosenberg, R.~O.\ 2003, Science, 299, 77 
\bibitem[Hamuy et al.(2002)]{2002AJ....124..417H} Hamuy, M., Maza, J., et 
al.\ 2002, \aj, 124, 417 
\bibitem[Hernandez et al.(2000)]{2000MNRAS.319..223H} Hernandez, M., 
Meikle, W.~P.~S., et al.\ 2000, \mnras, 319, 223 
\bibitem[Hicken et al.(2009)]{2009ApJ...700..331H} Hicken, M., Challis, P., 
et al.\ 2009, \apj, 700, 331 
\bibitem[H{\"o}flich et al.(1995)]{1995ApJ...444..831H} H{\"o}flich, P., Khokhlov, 
A.~M., \& Wheeler, J.~C.\ 1995, \apj, 444, 831 
\bibitem[H{\"o}flich et al.(2002)]{2002ApJ...568..791H} H{\"o}flich, P., 
Gerardy, C.~L., Fesen, R.~A., \& Sakai, S.\ 2002, \apj, 568, 791 
\bibitem[Horesh et al.(2012)]{2012ApJ...746...21H} Horesh, A., Kulkarni, 
S.~R., et al.\ 2012, \apj, 746, 21 
\bibitem[Hoyle 
\& Fowler(1960)]{1960ApJ...132..565H} Hoyle, F., \& Fowler, W.~A.\ 1960, \apj, 132, 565 
\bibitem[Hsiao et al.(2007)]{2007ApJ...663.1187H} Hsiao, E.~Y., Conley, A., 
et al.\ 2007, \apj, 663, 1187 
\bibitem[Hsiao(2009)]{2009PhDT.......228H} Hsiao, Y.~C.~E.\ 2009, 
Ph.D.~Thesis, University of Victoria
\bibitem[Iwamoto et al.(1999)]{1999ApJS..125..439I} Iwamoto, K., Brachwitz, 
F., et al.\ 1999, \apjs, 125, 439 
\bibitem[Jackson et al.(2010)]{2010ApJ...720...99J} Jackson, A.~P., Calder, 
A.~C., et al.\ 2010, \apj, 720, 99 
\bibitem[Jha et al.(1999)]{1999ApJS..125...73J} Jha, S., Garnavich, P.~M., 
et al.\ 1999, \apjs, 125, 73 
\bibitem[Kasen et al.(2009)]{2009Natur.460..869K} Kasen, D., R{\"o}pke, 
F.~K., \& Woosley, S.~E.\ 2009, \nat, 460, 869 
\bibitem[Kasen(2006)]{2006ApJ...649..939K} Kasen, D.\ 2006, \apj, 649, 939 
\bibitem[Kattner et al.(2012)]{2012PASP..124..114K} Kattner, S., Leonard, 
D.~C., et al.\ 2012, \pasp, 124, 114 
\bibitem[Kelly et al.(2010)]{2010ApJ...715..743K} Kelly, P.~L., Hicken, M., 
Burke, D.~L., Mandel, K.~S., \& Kirshner, R.~P.\ 2010, \apj, 715, 743 
\bibitem[Khokhlov(1991)]{1991A&A...245..114K} Khokhlov, A.~M.\ 1991, \aap, 245, 114 
\bibitem[Khokhlov et al.(1997)]{1997ApJ...478..678K} Khokhlov, A.~M., Oran, 
E.~S., \& Wheeler, J.~C.\ 1997, \apj, 478, 678 
\bibitem[Kirshner et al.(1973)]{1973ApJ...180L..97K} Kirshner, R.~P., 
Willner, S.~P., Becklin, E.~E., Neugebauer, G., 
\& Oke, J.~B.\ 1973, \apjl, 180, L97 
\bibitem[Krisciunas et al.(2007)]{2007AJ....133...58K} Krisciunas, K., 
Garnavich, P.~M., et al.\ 2007, \aj, 133, 58 
\bibitem[Krisciunas et al.(2009)]{2009AJ....138.1584K} Krisciunas, K., 
Marion, G.~H., et al.\ 2009, \aj, 138, 1584 
\bibitem[Krisciunas et al.(2004)]{2004ApJ...602L..81K} Krisciunas, K., 
Phillips, M.~M., \& Suntzeff, N.~B.\ 2004, \apjl, 602, L81 
\bibitem[Kromer 
\& Sim(2009)]{2009MNRAS.398.1809K} Kromer, M., \& Sim, S.~A.\ 2009, \mnras, 398, 1809 
\bibitem[Lampeitl et al.(2010)]{2010ApJ...722..566L} Lampeitl, H., Smith, 
M., et al.\ 2010, \apj, 722, 566 
\bibitem[Li et al.(2011)]{2011Natur.480..348L} Li, W., Bloom, J.~S., et 
al.\ 2011, \nat, 480, 348 
\bibitem[Lynch et al.(1990)]{1990AJ....100..223L} Lynch, D.~K., Rudy, 
R.~J., et al.\ 1990, \aj, 100, 223 
\bibitem[Mandel et al.(2011)]{2011ApJ...731..120M} Mandel, K.~S., Narayan, 
G., \& Kirshner, R.~P.\ 2011, \apj, 731, 120 
\bibitem[Mandel et al.(2009)]{2009ApJ...704..629M} Mandel, K.~S., 
Wood-Vasey, W.~M., Friedman, A.~S., 
\& Kirshner, R.~P.\ 2009, \apj, 704, 629 
\bibitem[Margutti et al.(2012)]{2012ApJ...751..134M} Margutti, R., 
Soderberg, A.~M., et al.\ 2012, \apj, 751, 134 
\bibitem[Marion et al.(2009)]{2009AJ....138..727M} Marion, G.~H., 
H{\"o}flich, P., et al.\ 2009, \aj, 138, 727 
\bibitem[Marion et al.(2003)]{2003ApJ...591..316M} Marion, G.~H., 
H{\"o}flich, P., Vacca, W.~D., \& Wheeler, J.~C.\ 2003, \apj, 591, 316 
\bibitem[Marion et al.(2001)]{2001RMxAC..10..190M} Marion, G.~H., 
H{\"o}flich, P., 
\& Wheeler, J.~C.\ 2001, Revista Mexicana de Astronomia y Astrofisica Conference Series, 10, 190 
\bibitem[Marion et al.(2006)]{2006ApJ...645.1392M} Marion, G.~H., 
H{\"o}flich, P., et al.\ 2006, \apj, 645, 1392 
\bibitem[Markwardt(2009)]{2009ASPC..411..251M} Markwardt, C.~B.\ 2009, 
Astronomical Data Analysis Software and Systems XVIII, 411, 251 
\bibitem[Matheson et al.(2012)]{2012ApJ...754...19M} Matheson, T., Joyce, 
R.~R., et al.\ 2012, \apj, 754, 19 
\bibitem[Meikle et al.(1996)]{1996MNRAS.281..263M} Meikle, W.~P.~S., 
Cumming, R.~J., et al.\ 1996, \mnras, 281, 263 
\bibitem[Niemeyer 
\& Woosley(1997)]{1997ApJ...475..740N} Niemeyer, J.~C., \& Woosley, S.~E.\ 1997, \apj, 475, 740 
\bibitem[Nomoto et 
al.(1976)]{1976Ap&SS..39L..37N} Nomoto, K., Sugimoto, D., \& Neo, S.\ 1976, \apss, 39, L37 
\bibitem[Nomoto et al.(1984)]{1984ApJ...286..644N} Nomoto, K., Thielemann, 
F.-K., \& Yokoi, K.\ 1984, \apj, 286, 644 
\bibitem[Nugent et al.(2002)]{2002PASP..114..803N} Nugent, P., Kim, A., 
\& Perlmutter, S.\ 2002, \pasp, 114, 803 
\bibitem[Nugent et al.(1995)]{1995ApJ...455L.147N} Nugent, P., Phillips, 
M., Baron, E., Branch, D., \& Hauschildt, P.\ 1995, \apjl, 455, L147 
\bibitem[Nugent et al.(2011)]{2011Natur.480..344N} Nugent, P.~E., Sullivan, 
M., et al.\ 2011, \nat, 480, 344 
\bibitem[Oke 
\& Sandage(1968)]{1968ApJ...154...21O} Oke, J.~B., \& Sandage, A.\ 1968, \apj, 154, 21 
\bibitem[Parrent et al.(2012)]{2012ApJ...752L..26P} Parrent, J.~T., Howell, 
D.~A., et al.\ 2012, \apjl, 752, L26 
\bibitem[Parrent et al.(2011)]{2011ApJ...732...30P} Parrent, J.~T., Thomas, 
R.~C., et al.\ 2011, \apj, 732, 30 
\bibitem[Patat et al.(2011)]{2011arXiv1112.0247P} Patat, F., Cordiner, 
M.~A., et al.\ 2011, arXiv:1112.0247 
\bibitem[Perlmutter et al.(1999)]{1999ApJ...517..565P} Perlmutter, S., 
Aldering, G., et al.\ 1999, \apj, 517, 565 
\bibitem[Phillips(1993)]{1993ApJ...413L.105P} Phillips, M.~M.\ 1993, \apjl, 
413, L105 
\bibitem[Pignata et al.(2008)]{2008MNRAS.388..971P} Pignata, G., Benetti, 
S., et al.\ 2008, \mnras, 388, 971 
\bibitem[Pinto 
\& Eastman(2000)]{2000ApJ...530..757P} Pinto, P.~A., \& Eastman, R.~G.\ 2000, \apj, 530, 757 
\bibitem[Plewa et al.(2004)]{2004ApJ...612L..37P} Plewa, T., Calder, A.~C., 
\& Lamb, D.~Q.\ 2004, \apjl, 612, L37 
\bibitem[Press et al.(1992)]{1992nrca.book.....P} Press, W.~H., Teukolsky, 
S.~A., Vetterling, W.~T., 
\& Flannery, B.~P.\ 1992, Cambridge: University Press, |c1992, 2nd ed.,  
\bibitem[R{\"o}pke et al.(2007)]{2007ApJ...668.1132R} R{\"o}pke, F.~K., 
Hillebrandt, W., et al.\ 2007, \apj, 668, 1132 
\bibitem[R{\"o}pke et al.(2012)]{2012ApJ...750L..19R} R{\"o}pke, F.~K., 
Kromer, M., et al.\ 2012, \apjl, 750, L19 
\bibitem[Rayner et al.(2003)]{2003PASP..115..362R} Rayner, J.~T., Toomey, 
D.~W., et al.\ 2003, \pasp, 115, 362 
\bibitem[Riess et al.(1998)]{1998AJ....116.1009R} Riess, A.~G., Filippenko, 
A.~V., et al.\ 1998, \aj, 116, 1009 
\bibitem[Silverman 
\& Filippenko(2012)]{2012MNRAS.425.1917S} Silverman, J.~M., \& Filippenko, A.~V.\ 2012, \mnras, 425, 1917 
\bibitem[Spyromilio et al.(1994)]{1994MNRAS.266L..17S} Spyromilio, J., 
Pinto, P.~A., \& Eastman, R.~G.\ 1994, \mnras, 266, L17 
\bibitem[Spyromilio et al.(1992)]{1992MNRAS.258P..53S} Spyromilio, J., 
Meikle, W.~P.~S., Allen, D.~A., \& Graham, J.~R.\ 1992, \mnras, 258, 53P 
\bibitem[Stanishev et 
al.(2007)]{2007A&A...469..645S} Stanishev, V., Goobar, A., et al.\ 2007, \aap, 469, 645 
\bibitem[Sullivan et al.(2010)]{2010MNRAS.406..782S} Sullivan, M., Conley, 
A., et al.\ 2010, \mnras, 406, 782 
\bibitem[Suzuki(2006)]{2006ApJS..163..110S} Suzuki, N.\ 2006, \apjs, 163, 
110 
\bibitem[Tanaka et al.(2008)]{2008ApJ...677..448T} Tanaka, M., Mazzali, 
P.~A., et al.\ 2008, \apj, 677, 448 
\bibitem[Thomas et al.(2007)]{2007ApJ...654L..53T} Thomas, R.~C., Aldering, 
G., et al.\ 2007, \apjl, 654, L53 
\bibitem[Thomas et al.(2011a)]{2011ApJ...743...27T} Thomas, R.~C., Aldering, 
G., et al.\ 2011a, \apj, 743, 27 
\bibitem[Thomas et al.(2011b)]{2011PASP..123..237T} Thomas, R.~C., Nugent, 
P.~E., \& Meza, J.~C.\ 2011b, \pasp, 123, 237 
\bibitem[Timmes et al.(2003)]{2003ApJ...590L..83T} Timmes, F.~X., Brown, 
E.~F., \& Truran, J.~W.\ 2003, \apjl, 590, L83 
\bibitem[Toth \& Szab{\'o}(2000)]{2000A&A...361...63T} Toth, I., \& Szab{\'o}, R.\ 2000, \aap, 361, 63
\bibitem[Vacca et al.(2003)]{2003PASP..115..389V} Vacca, W.~D., Cushing, 
M.~C., \& Rayner, J.~T.\ 2003, \pasp, 115, 389 
\bibitem[Vink{\'o} et 
al.(2012)]{2012A&A...546A..12V} Vink{\'o}, J., S{\'a}rneczky, K., et al.\ 2012, \aap, 546, A12 
\bibitem[Wheeler(2012)]{2012ApJ...758..123W} Wheeler, J.~C.\ 2012, \apj, 
758, 123 
\bibitem[Wheeler et al.(1998)]{1998ApJ...496..908W} Wheeler, J.~C., 
Hoeflich, P., Harkness, R.~P., \& Spyromilio, J.\ 1998, \apj, 496, 908 
\bibitem[Wood-Vasey et al.(2008)]{2008ApJ...689..377W} Wood-Vasey, W.~M., 
Friedman, A.~S., et al.\ 2008, \apj, 689, 377 
\bibitem[Woosley et al.(2011)]{2011ApJ...734...37W} Woosley, S.~E., 
Kerstein, A.~R., \& Aspden, A.~J.\ 2011, \apj, 734, 37 
\bibitem[Woosley et al.(2009)]{2009ApJ...704..255W} Woosley, S.~E., 
Kerstein, A.~R., Sankaran, V., Aspden, A.~J.,  
R{\"o}pke, F.~K.\ 2009, \apj, 704, 255 
\bibitem[Yamaoka et al.(1992)]{1992ApJ...393L..55Y} Yamaoka, H., Nomoto, 
K., Shigeyama, T., \& Thielemann, F.-K.\ 1992, \apjl, 393, L55 
\end{thebibliography}
\end{document}